\documentclass{article}

\usepackage[final]{neurips_data_2024}

\usepackage[utf8]{inputenc} 
\usepackage[T1]{fontenc}    
\usepackage{hyperref}       
\usepackage{url}            
\usepackage{booktabs}       
\usepackage{amsfonts}       
\usepackage{nicefrac}       
\usepackage{microtype}      
\usepackage[table]{xcolor}         
\usepackage{subfigure}

\newcommand{\model}[1]{CURE4Rec}
\usepackage{graphicx}
\usepackage{enumitem}
\usepackage{multicol}
\usepackage{multirow}
\usepackage{wrapfig}

\title{CURE4Rec: A Benchmark for Recommendation Unlearning with Deeper Influence}

\author{%
  Chaochao Chen$^1$,
  Jiaming Zhang$^1$,
  Yizhao Zhang$^1$,
  Li Zhang$^1$,
  Lingjuan Lyu$^2$,\\
  \textbf{Yuyuan Li$^{3, 1}$\thanks{Corresponding author} ,
  Biao Gong$^3$,
  Chenggang Yan$^3$
  }\\
  $^1$Zhejiang University, $^2$Sony AI, $^3$Hangzhou Dianzi University\\
  \\[1.5ex]
  \small{\texttt{\{zjuccc, 22321350, 22221337\}@zju.edu.cn, 
  zhanglizl80@gmail.com, lingjuan.lv@sony.com}}\\
  \small{\texttt{y2li@hdu.edu.cn, a.biao.gong@gmail.com, cgyan@hdu.edu.cn}} \\
}

\begin{document}

\maketitle

\begin{abstract}
With increasing privacy concerns in artificial intelligence, regulations have mandated the \textit{right to be forgotten}, granting individuals the right to withdraw their data from models. Machine unlearning has emerged as a potential solution to enable selective forgetting in models, particularly in recommender systems where historical data contains sensitive user information. Despite recent advances in recommendation unlearning, evaluating unlearning methods comprehensively remains challenging due to the absence of a unified evaluation framework and overlooked aspects of deeper influence, e.g., fairness.
To address these gaps, we propose \model{}, the first comprehensive benchmark for recommendation unlearning evaluation.
\model{} covers four aspects, i.e., unlearning Completeness, recommendation Utility, unleaRning efficiency, and recommendation fairnEss, under three data selection strategies, i.e., core data, edge data, and random data.
Specifically, we consider the deeper influence of unlearning on recommendation fairness and robustness towards data with varying impact levels.
We construct multiple datasets with \model{} evaluation and conduct extensive experiments on existing recommendation unlearning methods.
Our code is released at \url{https://github.com/xiye7lai/CURE4Rec}.
\end{abstract}

\section{Introduction}

Over the past few years, growing concerns over information abundance and data leakage have intensified the focus on privacy preservation within artificial intelligence.
Regulations such as the General Data Protection Regulation (GDPR)~\citep{euro2018gdpr}, the California Consumer Privacy Act~\citep{pardau2018california}and the Delete Act~\citep{cal2023del} grant individuals the \textit{right to be forgotten}, requiring the deletion of personal data used in information systems.
Nowadays, the ubiquitous application of machine learning models in information systems poses potential risks for memorizing training data~\citep{fredrikson2015model}. Consequently, the aforementioned regulations also require forgetting the associated data memory within the trained models, giving rise to the concept of machine unlearning. Recently, machine unlearning has gained increasing popularity in computer vision~\citep{bourtoule2021machine,gupta2021adaptive}, natural language processing~\citep{chen2023unlearn, eldan2023s}, and recommender systems~\citep{chen2022recommendation,li2023ultrare,li2023selective}.
As recommender systems typically rely on historical interaction data to extract user preferences, the recommendation model inherently contains sensitive user information. Therefore, there is a crucial need for unlearning to preserve privacy. The task of machine unlearning in recommender systems is termed as recommendation unlearning.

While machine unlearning has demonstrated significant potential in preserving user privacy, conducting a comprehensive evaluation of unlearning methods continues to pose difficulties.
Various unlearning methods employ distinct evaluation metrics, yet a universally applicable evaluation framework remains absent.
Specifically, existing evaluation methods predominantly focus on the unlearning completeness, unlearning efficiency, and its impact on model utility, overlooking the deeper influence of model properties. 

In this paper, we identify two overlooked aspects of deeper influence. \textit{Firstly}, fairness is a crucial consideration for recommendations~\citep{wang2023survey}, but is often neglected in unlearning evaluations. Ensuring fair recommendation outcomes can avoid user discrimination and enrich the recommendation platform's understanding of user preferences. Existing studies demonstrate that unlearning can affect the fairness of models~\citep{oesterling2024fair}.
\textit{Secondly}, existing evaluation methods neglect the influence of various unlearning sets, randomly selecting data for unlearning. Distinct unlearning sets, however, can result in significantly different impacts on model performance~\citep{fan2024challenging}. Performing comprehensive evaluations on different unlearning data contributes to understanding the robustness of unlearning methods.
 
\begin{figure}
    \centering
    \includegraphics[width=0.85\linewidth]{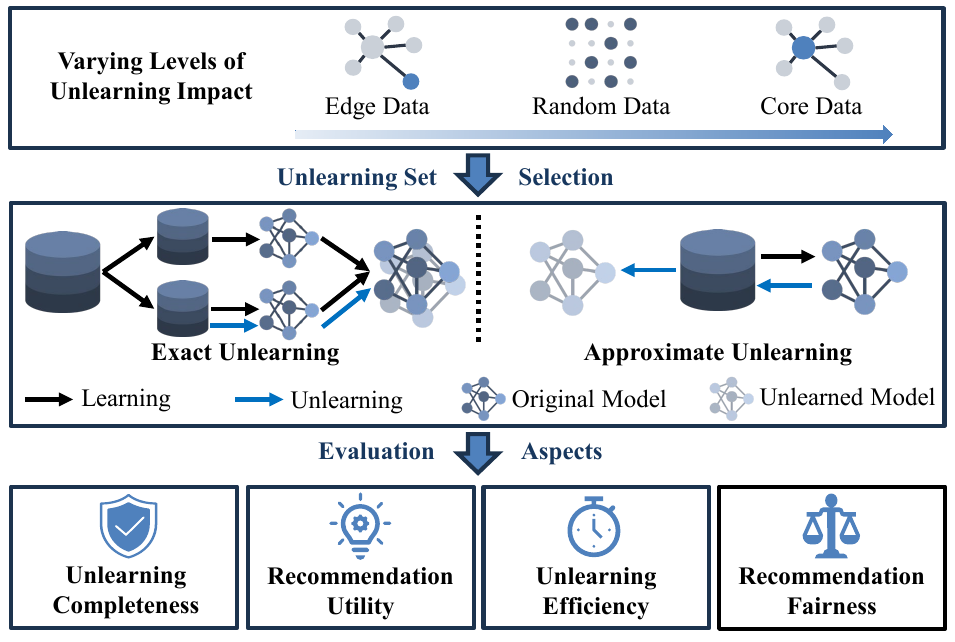}
    \caption{An illustration of \model{}, a comprehensive benchmark tailored for evaluating recommendation unlearning methods. \model{} evaluates unlearning methods using data with varying levels of unlearning impact on four aspects, i.e., unlearning completeness, recommendation utility, unlearning efficiency, and recommendation fairness.}
    \label{fig:frame}
\end{figure}

To address these issues, we introduce \model{}, a comprehensive benchmark specifically designed to evaluate recommendation unlearning methods. As shown in Figure~\ref{fig:frame}, \model{}'s evaluation encompasses four aspects, i.e., unlearning completeness, recommendation utility, unlearning efficiency, and recommendation fairness. Additionally, each aspect is investigated with three data selection strategies, i.e., core data, edge data, and random data. This triadic breakdown tests to reflect the robustness of recommendation unlearning methods towards different unlearning sets.
The main contributions of this work are summarized as follows:
\begin{itemize}
\item We introduce \model{}, a comprehensive benchmark tailored for evaluating recommendation unlearning methods. \model{} enables evaluation across multiple aspects, including unlearning completeness, recommendation utility, unlearning efficiency, and recommendation fairness.
\item To the best of our knowledge, we are the first to investigate the impact of unlearning on recommendation fairness, introducing fairness evaluation to comprehensively grasp its impact and proposing additional requirements to consider for further research.
\item We further examine the impact of different unlearning sets. Based on the level of collaboration, we select core data, edge data, and random data to construct unlearning sets respectively, aiming to thoroughly explore the impact towards unlearning completeness, recommendation utility, unlearning efficiency, and recommendation fairness.
\item We offer multiple datasets tailored for evaluation using our \model{}. Furthermore, we conduct extensive experiments across existing recommendation unlearning methods and report their performance (please refer to Figure~\ref{fig:overview} for an overview of our results).
\end{itemize}

\section{Related Work}
\subsection{Machine Unlearning}
Machine unlearning aims to eliminate the memory of specific data, serving purposes such as privacy protection~\citep{liu2022backdoor} and erasing data biases~\citep{sattigeri2022fair, chen2024fast}. According to the level of unlearning completeness, existing machine unlearning methods can be categorized into two approaches, i.e., exact unlearning and approximate unlearning. 

\textit{Exact Unlearning (EU)} aims to completely eliminate the influence of target data on the model. 
The most straightforward method of exact unlearning is retraining the model from scratch on the updated dataset (removing the target data), but this method incurs a significant computational time cost. To mitigate this cost, existing EU methods revamp the training process via ensemble learning, which limits the retraining cost to sub-datasets or sub-models~\citep{bourtoule2021machine,yan2022arcane}. 

\textit{Approximate Unlearning (AU)} achieves unlearning through direct parameter manipulation, avoiding the significant time cost of retraining. Most AU methods utilize gradients or influence function to estimate the influence of target data and subsequently remove it from models~\citep{sekhari2021remember,wu2022puma,mehta2022deep}. 
Alternatively, other methods directly prune or dampen model parameters to achieve unlearning~\citep{wang2022federated,foster2024fast}.


\subsection{Recommendation Unlearning}
Recommendation unlearning aims to eliminate the influence of target data within the recommender system~\citep{li2024survey}. 
A naive approach to achieve recommendation unlearning is through the direct application of the classic unlearning method, i.e., SISA~\citep{bourtoule2021machine}.
Due to the collaborative characteristics of recommendation data, tailored methods have been proposed to improve SISA for recommendation unlearning, e.g., RecEraser~\citep{chen2022recommendation} and UltraRE~\citep{li2023ultrare}.
In addition to EU methods mentioned above, AU method also enters the scene, utilizing refined influence functions to enable recommendation unlearning~\citep{li2023selective, zhang2023recommendation}. 
Note that this paper focuses on investigating the model-agnostic approaches. 
Other approaches focus on specific models, e.g., sequential recommendation~\citep{ye2023sequence}, session-based recommendation~\citep{xin2024effectiveness}, and large language model-based recommendation~\citep{wang2024towards, hu2024exact}.
Note that this paper focuses solely on recommendation unlearning of training data. An alternative line of research, known as attribute unlearning, explores the unlearning of latent user attributes in recommender systems.~\citep{ganhor2022unlearning, li2023making, chen2024post}.

\subsection{Machine Unlearning Benchmarks}
Emerging research has pioneered early investigation into unlearning benchmarks, focusing on image classification~\citep{choi2023towards}, large language models~\citep{maini2024tofu, li2024wmdp, jin2024rwku}, and diffusion models~\citep{zhang2024unlearncanvas}.
By proposing new datasets or modifying existing ones, these investigations design depth evaluation metrics within their corresponding domains. 
However, these benchmarks leave unexplored deeper influence of unlearning on model properties, i.e., fairness and robustness.
This exploration is crucial for recommender systems, as alternations in the performance of recommendation models immediately affect recommendation lists, eventually influencing use experience.
To the best of our knowledge, we are the first to introduce a recommendation unlearning benchmark, and comprehensively explore the deeper influence of unlearning on recommendation fairness and robustness.

\section{\model{}}
In this section, we first recall the process of recommendation unlearning, outlining the necessary inputs for evaluations. Then, we introduce evaluation aspects of our proposed \model{}, detailing specific metrics for each aspect. Finally, we present the strategy for unlearning set selection.

\subsection{Recommendation Unlearning}
The entire process of recommendation unlearning consists of four stages: I) completing learning process to generate the original model; II) determining the unlearning set, i.e., the unlearning target, which is a subset of training data; III) conducting unlearning process based on the original model to produce the unlearned model; and IV) evaluating the unlearned model.
To ensure reliable evaluation, we evaluate unlearning methods using identical training and testing data, employing the same learning process to generate the same original model. This ensures that all unlearning methods start from the same baseline in stage I. To investigate unlearning robustness, we select three types of unlearning sets in stage II (Section~\ref{sec:select}). In stage IV, \model{}'s evaluation includes the four aspects (Section~\ref{sec:aspect}).

In the context of recommendation, unlearning targets may vary among users, items, and user-item interactions. Commonly, recommendation unlearning scenarios focus on user-wise unlearning~\citep{li2023ultrare}. 
Thus, our benchmark primarily investigates the user-wise unlearning scenarios.

\subsection{Evaluation Aspects}
\label{sec:aspect}

\paragraph{Unlearning Completeness.}
Unlearning completeness stands as the primary goal and fundamental requirement of recommendation unlearning.
Exact unlearning methods inherently guarantee unlearning completeness by retraining, which is the only authorized approach~\citep{thudi2022necessity}.
On the other hand, approximate unlearning methods, lacking the ability to achieve authorized unlearning, often require the demonstration of unlearning completeness through theoretical proofs or empirical studies.
Therefore, following the completeness evaluation of approximate unlearning in previous studies~\citep{graves2021amnesiac,ma2022learn,li2023selective,kurmanji2024towards}, we evaluate unlearning completeness of recommendation unlearning based on the attacking performance of Membership Inference Oracle (MIO). 

MIO follows the standard membership inference procedure to evaluate unlearning completeness in image classification task~\citep{graves2021amnesiac,ma2022learn}.
In the context of recommendation, we concatenate user embeddings with the average item embeddings of their respective interacted items as the data features, and the probability of being in the training set as the data label.
Please refer to Section~\ref{sec:mio} for more training details.
To evaluate unlearning completeness, we query MIO with the unlearned data points. Ideally, MIO outputs 1 (indicating presence in the training set) for the original model and 0 (indicating absence from the training set) for the unlearned model. 
Since exact unlearning methods guarantee complete unlearning, we only evaluate the completeness of approximate unlearning methods.

\paragraph{Recommendation Utility.} Recommendation unlearning aims to erase the memory of target data within recommender systems without causing harm to the knowledge acquired from the remaining data. Thus, preserving the recommendation utility of the remaining data is another important goal of unlearning.
To investigate the impact of unlearning on model utility, we employ two widely used metrics, i.e., Normalized Discounted Cumulative Gain (NDCG) and Hit Ratio (HR), to evaluate the recommendation performance of the unlearned model on the testing set.
For both metrics, we truncate the ranked list to 20 items.

\paragraph{Unlearning Efficiency.} 
Retraining from scratch represents the gold standard in unlearning, but its practical implementation carries a prohibitive computational overhead. 
Recommender systems encompass hundreds of thousands of users, generating a large amount of unlearning requests. Therefore, improving unlearning efficiency is a crucial goal of recommendation unlearning. 
We measure unlearning efficiency by the total runtime of the entire unlearning process, i.e., stage III. Note that we enable parallel training for exact unlearning.

\paragraph{Recommendation Fairness.} 
Previous research has demonstrated that unlearning affects deeper model properties such as fairness~\citep{oesterling2024fair}. Mitigating the negative impact of unlearning is also an important requirement of unlearning.
In this paper, we evaluate the performance fairness of recommendation unlearning from the following two perspectives: i) the fairness between active and inactive groups (A-IGF), and ii) the fairness among different shards (shardGF), as exact unlearning methods divide the datasets into multiple shards.

For A-IGF, we follow the representative user-oriented group fairness research in recommendation~\citep{li2021user}. 
Based on the number of interactions, we classify the top 5\% of users as the active group and the remaining 95\% users as the inactive group. Active and inactive users are selected outside the unlearning set, because we aim to investigate the impact on the remainder users.
Then we compute the difference of the average recommendation utility, i.e., NDCG@20, between active and inactive groups to represent A-IGF.
For shardGF, we report the variance of recommendation utility among all shards to compare the shard-level fairness~\citep{rastegarpanah2019fighting}. 
Note that we do not compute shardGF for approximate unlearning, because these methods do not involve sharding.

\subsection{Unlearning Set Selection}\label{sec:select}
Existing evaluation methods typically select data randomly for the unlearning set. However, previous studies have shown that i) poisoned data can be constructed to make it hard to unlearn~\citep{marchant2022hard}, and ii) different data points have varying difficulty of unlearning~\citep{fan2024challenging}. 
Motivated by these findings, in this paper, we explore the impact of using varying unlearning sets, which can also reflect the robustness of unlearning. 

To significantly demonstrate this impact, we adopt a model-agnostic selection strategy to create three types of unlearning sets: core data (which impacts many other data points), edge data (with minimal impact on others), and random data.
Specifically, we regard the user-item interactions as a non-weighted bipartite graph, where users and items are represented as nodes, and an edge connects them if there is an interaction. 
%
%
Existing research suggests that a node's importance correlates strongly with its centrality in a graph~\citep{haveliwala2002topic,li2012har,park2019estimating}.
In the context of recommendation, centrality is associated with collaborations, manifested as neighbors in a graph. Thus, we define the importance of a node $x$ as follows: 
\begin{equation}
    I(x)=c(x) \cdot \frac{\sum_{y \in N(x)} c(y)}{\left | N(x) \right | },
\end{equation}
where $c(x)$ denotes the centrality of node $x$, and $N(x)$ denotes the number of neighbors of node $x$.
Due to the collaborative characteristic of recommendation data, we use the degree of node, i.e., the number of first-order neighbors, to compute centrality.
Finally, we rank all nodes based on $I(x)$ to select the core data and edge data.

\section{Experimental Setup}

\subsection{Datasets}\label{sec:dataset}

We conduct experiments on three real-world datasets widely used in recommendation.
\textbf{ML-100K}\footnote{https://grouplens.org/datasets/movielens/}: The MovieLens dataset is one of the most extensively utilized datasets in recommender system research. MovieLens 100k contains 100 thousand individual ratings.
\textbf{ML-1M}: MovieLens 1M contains 1 million ratings.
\textbf{ADM}\footnote{http://jmcauley.ucsd.edu/data/amazon/}: The Amazon dataset comprises multiple subsets categorized according to different types of Amazon products. One of these subsets, known as the Amazon Digital Music (ADM) dataset, includes ratings of digital music.
Following the widely used pre-processing procedure~\citep{he2017neural,wang2019neural,he2020lightgcn}, we convert ratings into implicit feedback.
The statistical details of these datasets are summarized in Table~\ref{tab:dataset}. To avoid extreme sparsity, we filter out the users and items that have less than 5 interactions. 
For each dataset, we randomly select 80\% ratings as the training set, 10\% ratings as the validation set, and the remaining as the test set. 
The unlearning ratio, i.e., the percentage of unlearning set within the training set, is initially set as 5\%. We also explore this ratio within a range of (5\%, 10\%, 15\%, 20\%) in Appendix~\ref{sec:ratio}.

\begin{table}[h]
\centering
\caption{Summary of datasets.}
\label{tab:dataset}
\begin{tabular}{lcccc}  
\toprule
Dataset & User \#   & Item \#   & Interactions \# & Sparsity \\
\midrule
ML-100K & 943 & 1,349 & 99,287 & 92.195\% \\
ML-1M  & 6,040 & 3,416  & 999,611 & 95.155\% \\
ADM  &  478,235 & 266,414   & 836,006  & 99.999\% \\
\bottomrule
\end{tabular}
\end{table}

\subsection{Recommendation Models}
Aligning with existing studies on recommendation unlearning~\citep{chen2022recommendation, li2023selective, li2023ultrare}, we use three representative recommendation models based on collaborative filtering for evaluation:
\begin{itemize}
\item \textbf{WMF}: Weighted Matrix Factorization(WMF)~\citep{chen2020efficient} is a non-sampling recommendation model that treats all missing interactions as 
negative interactions and assigns them with uniform weights. 
\item \textbf{BPR}: Bayesian Personalized Ranking~\citep{rendle2012bpr} is a widely used recommendation model that uses a Bayesian personalized ranking objective function to optimize matrix factorization.
\item \textbf{LightGCN}: LightGCN~\citep{he2020lightgcn} is the state-of-the-art collaborative filtering model, which improves recommendation performance by simplifying graph neural networks.
\end{itemize}

\subsection{Unlearning Methods}
We consider the following recommendation unlearning methods, including both EU and AU approaches (note that we set the number of shards to 10 for EU and explore other values in Section~\ref{sec:shard}):
\begin{itemize}
\item \textbf{Retrain}: Retraining from scratch is the goal standard unlearning method.
\item \textbf{SISA}: SISA~\citep{bourtoule2021machine} stands as the classic algorithm for machine unlearning, adaptable to various scenarios, including recommender systems. 

\item \textbf{RecEraser}: RecEraser~\citep{chen2022recommendation} is specifically designed for recommendation unlearning, which modifies SISA to boost performance in recommendation tasks. 
\item \textbf{UltraRE}: UltraRE~\citep{li2023ultrare} enhances RecEraser for recommendation tasks by modifying two key stages, i.e., division and aggregation.
\item \textbf{SCIF}: SCIF~\citep{li2023selective} is the first approximate unlearning method in recommendation systems, employing influence functions tailored for recommendation tasks.
\end{itemize}

\subsection{Parameters Settings}\label{sec:para}
In the training phase of original models, we randomly sample 4 negative items for each observed interaction following~\citep{he2017neural}.
%
%
In the case of model-specific hyper-parameters, we tune them in the ranges suggested by their original papers. In detail, the batch size is set to 512, the learning rate is set to 0.01, the embedding size is set to 32. 
The maximum number of epochs is set to 500. 
The early stopping strategy is adopted in our experiments, which terminates the training when NDCG@20 on the validation set does not increase for 5 successive epochs.

\subsection{MIO Training Details}\label{sec:mio}

%
Following~\citep{li2023selective}, we adopt an ideal concept, i.e., Membership Inference Oracle (MIO), to evaluate unlearning completeness. Specifically, We implement an approximated MIO via a basic three-layer (64, 16, 4) neural network with ReLu and Softmax as activation functions for hidden layers and the output layer respectively. We train the MIO via stochastic gradient descent with 100 epochs and a learning rate of 0.001. The MIO outputs the probability of the queried data point being in the training set. To evaluate the unlearning completeness, we query MIO with the unlearned data points. Ideally, MIO outputs 1 (being in the training set) for the original model while outputs 0 (not being in the training set) for the unlearned model.

\subsection{Hardware Information}\label{sec:hard}
We run all experiments on the same Ubuntu 20.04 LTS System server with 48-core CPU, 256GB RAM and NVIDIA GeForce RTX 3090 GPU.

\begin{wrapfigure}[20]{r}{0.48\textwidth}
    \vspace{-10pt}
    \centering
    \includegraphics[width=\linewidth]{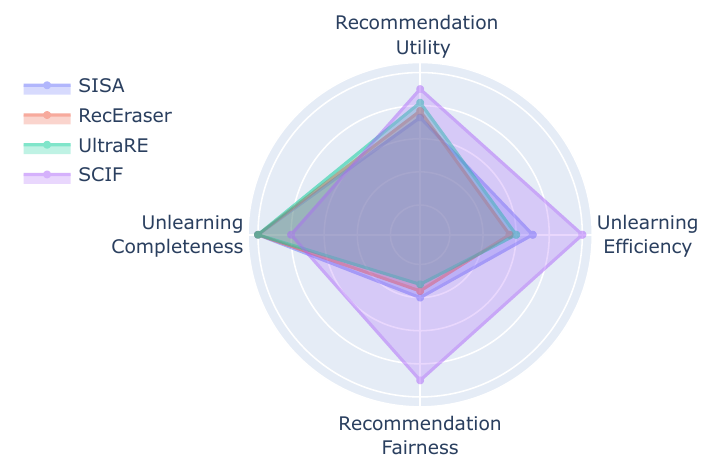}
    \vspace{-10pt}
    \caption{A visualized evaluation overview of recommendation unlearning methods in four aspects ($\uparrow$), where the result is the normalized average outcome obtained across all models and datasets, using random data as the unlearning set. 
    The recommendation fairness is measured by A-IGF (fairness between active and inactive users).
    The higher values represent better performance.
    }
    \label{fig:overview}
\end{wrapfigure}

\section{Results and Discussion}
In this section, we report and analyze the results regarding four evaluation aspects under three selections of unlearning sets. We present a visualized overview of compared recommendation unlearning methods in Figure~\ref{fig:overview}.
We observe that apart from unlearning completeness, the AU method (SCIF) demonstrates a significant advantage over EU methods (SISA, RecEraser, and UltraRE), particularly in terms of unlearning efficiency and recommendation fairness.
However, it is essential to highlight that unlearning completeness is the primary goal of unlearning. 
EU methods inherently achieve the highest level of completeness, whereas SCIF can only achieve weak unlearning.

\begin{table*}[t]
\caption{Results in terms of unlearning completeness (MIO accuracy - approaching 0.5), recommendation utility (NDCG and HR $\uparrow$), and recommendation fairness (A-IGF - approaching Retrain) for the approximate recommendation unlearning method, where \texttt{Learn} denotes the results before unlearning.
Core, random, and edge respectively refer to the selection of the unlearning sets as core data, random data, and edge data.}
\label{tab:au}
\centering
\resizebox{\textwidth}{!}{
\begin{tabular}{cl|rrrr|rrrr|rrrr}
    \toprule
    \multicolumn{2}{c}{\multirow{2}[4]{*}{}} & \multicolumn{4}{c}{ML-100K}   & \multicolumn{4}{c}{ML-1M}     & \multicolumn{4}{c}{ADM} \\
\cmidrule(r){3-6} \cmidrule(r){7-10} \cmidrule(r){11-14}  \multicolumn{2}{c}{} & NDCG@20 & HR@20 & MIO & A-IGF & NDCG@20 & HR@20 & MIO & A-IGF & NDCG@20 & HR@20 & MIO & A-IGF \\
    \midrule
    \texttt{Learn} &       & 0.3215  & 0.3415  & 0.722  & -0.0450  & 0.2144 & 0.2112 & 0.741 & -0.042 & 0.0277 & 0.0578 & 0.756 & 0.0167 \\
    \midrule
    \multirow{3}[2]{*}{Retrain} & Core  & 0.3187  & 0.3295  & 0.540  & -0.0184  & 0.2196  & 0.2174  & 0.544 & -0.0188  & 0.0221  & 0.0446  & 0.555 & 0.0053  \\
          & Random & 0.2872  & 0.3353  & 0.538  & -0.0403  & 0.2124  & 0.2108  & 0.547 & -0.0507  & 0.0252  & 0.0519  & 0.556 & 0.0141  \\
          & Edge  & 0.3091  & 0.3140  & 0.536  & -0.0430  & 0.2148  & 0.2051  & 0.546 & -0.0518  & 0.0272  & 0.0554  & 0.556 & 0.0164  \\
    \midrule
    \multirow{3}[2]{*}{SCIF} & Core  & 0.2483 & 0.2382 & 0.561  & -0.0322 & 0.1865 & 0.1629 & 0.569 & -0.0213 & 0.0194 & 0.0398 & 0.571 & 0.0094 \\
          & Random & 0.2699  & 0.2617  & 0.563  & -0.0268  & 0.1922  & 0.1785  & 0.571 & -0.0311  & 0.0227  & 0.0461  & 0.575 & 0.0106  \\
          & Edge  & 0.2894 & 0.3012 & 0.601  & -0.0375 & 0.2031 & 0.1811 & 0.623 & -0.0191 & 0.0245 & 0.0502 & 0.579 & 0.0103 \\
    \bottomrule
    \end{tabular}
}
\end{table*}

\subsection{Unlearning Completeness} \label{sec:completeness}
To evaluate the completeness of AU methods, we report the accuracy of MIO in Table~\ref{tab:au}, where the recommendation model is WMF. Due to the space limit, we report the results of other models in Appendix~\ref{sec:comp_app}.
Compared the result of SCIF with the performance before unlearning and Retrain after unlearning, we observe that i) both SCIF and Retrain significantly decrease the MIO accuracy, indicating their effectiveness in unlearning; ii) although not significant, there is still a marginal gap between SCIF and Retrain (ground truth), i.e., 4.1\% higher accuracy than Retrain on average; and iii) SCIF particularly performance worse on edge data compared to other data types. This discrepancy may be attributed to imprecise influence estimation for this specific data category.

\begin{table*}[t]
\caption{Results in terms of recommendation utility for exact recommendation unlearning methods. 
}
\label{tab:overview}
\centering
\resizebox{\textwidth}{!}{
\begin{tabular}{cl|rrr|rrr|rrr|rrr}
    \toprule
    \multicolumn{2}{c}{\multirow{2}[4]{*}{ML-100K}} & \multicolumn{3}{c}{Retrain} & \multicolumn{3}{c}{SISA} & \multicolumn{3}{c}{RecEraser} & \multicolumn{3}{c}{UltraRE} \\
\cmidrule(r){3-5} \cmidrule(r){6-8} \cmidrule(r){9-11} \cmidrule(r){12-14}     \multicolumn{2}{c}{} & \multicolumn{1}{l}{Core} & \multicolumn{1}{l}{Random} & \multicolumn{1}{l}{Edge} & \multicolumn{1}{l}{Core} & \multicolumn{1}{l}{Random} & \multicolumn{1}{l}{Edge} & \multicolumn{1}{l}{Core} & \multicolumn{1}{l}{Random} & \multicolumn{1}{l}{Edge} & \multicolumn{1}{l}{Core} & \multicolumn{1}{l}{Random} & \multicolumn{1}{l}{Edge} \\
    \midrule
    \multirow{2}[2]{*}{WMF} & NDCG@20 & 0.3187  & 0.2872  & 0.3091  & 0.2096  & 0.2092  & 0.2041  & 0.2285  & 0.2208  & 0.2109  & 0.2303  & 0.2354  & 0.2149  \\
          & HR@20 & 0.3295  & 0.3353  & 0.3140  & 0.2094  & 0.2049  & 0.1892  & 0.2218  & 0.2142  & 0.1979  & 0.2267  & 0.2282  & 0.2027  \\
    \midrule
    \multirow{2}[2]{*}{BPR} & NDCG@20 & 0.3111  & 0.3003  & 0.3043  & 0.2244  & 0.2324  & 0.2298  & 0.2614  & 0.2615  & 0.2694  & 0.2708  & 0.2764  & 0.2743  \\
          & HR@20 & 0.3151  & 0.3028  & 0.2987  & 0.2203  & 0.2259  & 0.2179  & 0.2724  & 0.2658  & 0.2620  & 0.2851  & 0.2813  & 0.2695  \\
    \midrule
    \multirow{2}[2]{*}{LightGCN} & NDCG@20 & 0.3175  & 0.3121  & 0.3101  & 0.1802  & 0.1932  & 0.1964  & 0.2856  & 0.2905  & 0.2886  & 0.2952  & 0.3069  & 0.3063  \\
          & HR@20 & 0.3250  & 0.3253  & 0.3244  & 0.1724  & 0.1907  & 0.1911  & 0.3053  & 0.3099  & 0.3121  & 0.3123  & 0.3201  & 0.3185 \\
    \midrule
    \midrule
    \multicolumn{2}{c}{\multirow{2}[4]{*}{ML-1M}} & \multicolumn{3}{c}{Retrain} & \multicolumn{3}{c}{SISA} & \multicolumn{3}{c}{RecEraser} & \multicolumn{3}{c}{UltraRE} \\
\cmidrule(r){3-5} \cmidrule(r){6-8} \cmidrule(r){9-11} \cmidrule(r){12-14}    \multicolumn{2}{c}{} & \multicolumn{1}{l}{Core} & \multicolumn{1}{l}{Random} & \multicolumn{1}{l}{Edge} & \multicolumn{1}{l}{Core} & \multicolumn{1}{l}{Random} & \multicolumn{1}{l}{Edge} & \multicolumn{1}{l}{Core} & \multicolumn{1}{l}{Random} & \multicolumn{1}{l}{Edge} & \multicolumn{1}{l}{Core} & \multicolumn{1}{l}{Random} & \multicolumn{1}{l}{Edge} \\
    \midrule
    \multirow{2}[2]{*}{WMF} & NDCG@20 & 0.2196  & 0.2124  & 0.2148  & 0.1780  & 0.1639  & 0.1714  & 0.1894  & 0.1796  & 0.1838  & 0.1926  & 0.1891  & 0.1970  \\
          & HR@20 & 0.2174  & 0.2108  & 0.2051  & 0.1612  & 0.1485  & 0.1493  & 0.1731  & 0.1592  & 0.1596  & 0.1747  & 0.1680  & 0.1717  \\
    \midrule
    \multirow{2}[2]{*}{BPR} & NDCG@20 & 0.2462  & 0.2319  & 0.2336  & 0.1545  & 0.1530  & 0.1628  & 0.1826  & 0.1660  & 0.1860  & 0.1828  & 0.1856  & 0.1913  \\
          & HR@20 & 0.2279  & 0.2162  & 0.2118  & 0.1353  & 0.1329  & 0.1367  & 0.1627  & 0.1450  & 0.1624  & 0.1652  & 0.1632  & 0.1651  \\
    \midrule
    \multirow{2}[2]{*}{LightGCN} & NDCG@20 & 0.2177  & 0.2108  & 0.2147  & 0.1504  & 0.1533  & 0.1642  & 0.1864  & 0.1863  & 0.1814  & 0.1969  & 0.1867  & 0.1806  \\
          & HR@20 &0.2138  & 0.2045  & 0.2186  & 0.1365  & 0.1323  & 0.1581  & 0.1825  & 0.1804  & 0.1818  & 0.1907  & 0.1855  & 0.1798  \\
    \midrule
    \midrule
    \multicolumn{2}{c}{\multirow{2}[4]{*}{ADM}} & \multicolumn{3}{c}{Retrain} & \multicolumn{3}{c}{SISA} & \multicolumn{3}{c}{RecEraser} & \multicolumn{3}{c}{UltraRE} \\
\cmidrule(r){3-5} \cmidrule(r){6-8} \cmidrule(r){9-11} \cmidrule(r){12-14}       \multicolumn{2}{c}{} & \multicolumn{1}{l}{Core} & \multicolumn{1}{l}{Random} & \multicolumn{1}{l}{Edge} & \multicolumn{1}{l}{Core} & \multicolumn{1}{l}{Random} & \multicolumn{1}{l}{Edge} & \multicolumn{1}{l}{Core} & \multicolumn{1}{l}{Random} & \multicolumn{1}{l}{Edge} & \multicolumn{1}{l}{Core} & \multicolumn{1}{l}{Random} & \multicolumn{1}{l}{Edge} \\
    \midrule
    \multirow{2}[2]{*}{WMF} & NDCG@20 & 0.3691  & 0.3566  & 0.3556  & 0.2720  & 0.2589  & 0.2515  & 0.3373  & 0.3256  & 0.3185  & 0.3420  & 0.3334  & 0.3347  \\
          & HR@20 & 0.4071  & 0.3822  & 0.3848  & 0.2617  & 0.2492  & 0.2471  & 0.3527  & 0.3467  & 0.3203  & 0.3689  & 0.3595  & 0.3501  \\
    \midrule
    \multirow{2}[2]{*}{BPR} & NDCG@20 & 0.3566  & 0.3453  & 0.3499  & 0.2806  & 0.2708  & 0.2757  & 0.3286  & 0.3295  & 0.3212  & 0.3325  & 0.3301  & 0.3314  \\
          & HR@20 & 0.3821  & 0.3628  & 0.3718  & 0.2745  & 0.2638  & 0.2611  & 0.3486  & 0.3406  & 0.3483  & 0.3541  & 0.3569  & 0.3608  \\
    \midrule
    \multirow{2}[2]{*}{LightGCN} & NDCG@20 & 0.0105  & 0.0106  & 0.0096  & 0.0075  & 0.0054  & 0.0048  & 0.0084  & 0.0085  & 0.0079  & 0.0097  & 0.0088  & 0.0086  \\
          & HR@20 &  0.0221  & 0.0234  & 0.0208  & 0.0157  & 0.0112  & 0.0103  & 0.0171  & 0.0176  & 0.0154  & 0.0191  & 0.0185  & 0.0183 \\
    \bottomrule
    \end{tabular}
}
\end{table*}

\subsection{Recommendation Utility}
We report the results in terms of recommendation utility for AU and EU in Tables~\ref{tab:au} and \ref{tab:overview}, respectively. 
In general, the AU method (SCIF) outperforms the EU methods (SISA, RecEraser, and UltraRE).
Employing the same unlearning set, RecEraser and UltraRE consistently outperform SISA across all datasets and models, with UltraRE generally surpassing RecEraser, aligning with previous research~\citep{li2023ultrare}.

For all EU methods, the recommendation utility of unlearning core users is generally higher than that of unlearning random-select or edge users. 
This is likely due to the removal of data from more interactive users, which typically contains a large amount of ratings. This enables the model to learn more effectively from the smaller amount of remaining training data. 
Compared with these EU methods, SCIF exhibits the highest recommendation utility, closely resembling that of Retrain.
However, SCIF suffers the most substantial performance decline when unlearning core users. 
This can be attributed to the increased number of interactions involved in calculating the influence function, leading to inaccurate influence estimation that negatively impacts the model utility.

\begin{table*}[t]
\caption{Results in terms of unlearning efficiency (running time in seconds $\downarrow$). 
}
\label{tab:time}
\centering
\resizebox{\textwidth}{!}{
\begin{tabular}{cl|rrr|rrr|rrr}
    \toprule
    \multicolumn{2}{c}{\multirow{2}[4]{*}{Time (s)}} & \multicolumn{3}{c}{ML-100K} & \multicolumn{3}{c}{ML-1M} & \multicolumn{3}{c}{ADM} \\
\cmidrule{3-11}    \multicolumn{2}{c}{} & \multicolumn{1}{l}{WMF} & \multicolumn{1}{l}{BPR} & \multicolumn{1}{l}{LightGCN} & \multicolumn{1}{l}{WMF} & \multicolumn{1}{l}{BPR} & \multicolumn{1}{l}{LightGCN} & \multicolumn{1}{l}{WMF} & \multicolumn{1}{l}{BPR} & \multicolumn{1}{l}{LightGCN} \\
    \midrule
    \multirow{3}[2]{*}{Retrain} & Core  & 4296  & 5238  & 4734  & 7748  & 9113  & 8645  & 3682  & 6998  & 5225  \\
          & Random & 4526  & 5494  & 5044  & 8693  & 9461  & 10324  & 3972  & 7127  & 5354  \\
          & Edge  & 4687  & 5527  & 5274  & 8006  & 9748  & 10497  & 4127  & 7351  & 6359  \\
    \midrule
    \multirow{3}[2]{*}{SISA} & Core  & 402   & 488   & 437   & 1160  & 1160  & 1523  & 669   & 1750  & 1009  \\
          & Random & 467   & 586   & 528   & 1256  & 1265  & 1605  & 717   & 1842  & 1246  \\
          & Edge  & 442   & 504   & 515   & 1280  & 1292  & 1659  & 751   & 1902  & 1077  \\
    \midrule
    \multirow{3}[2]{*}{RecEraser} & Core  & 463   & 582   & 561   & 1533  & 1568  & 1846  & 865   & 1892  & 1106  \\
          & Random & 476   & 693   & 656   & 1654  & 1660  & 1952  & 912   & 1945  & 1490  \\
          & Edge  & 489   & 659   & 617   & 1736  & 1819  & 1964  & 965   & 2032  & 1190  \\
    \midrule
    \multirow{3}[2]{*}{UltraRE} & Core  & 457   & 591   & 559   & 1507  & 1493  & 1667  & 819   & 1810  & 1057  \\
          & Random & 482   & 618   & 645   & 1595  & 1550  & 1834  & 901   & 1862  & 1283  \\
          & Edge  & 466   & 518   & 666   & 1781  & 1791  & 1955  & 923   & 1904  & 1368  \\
    \midrule
    \multirow{3}[2]{*}{SCIF} & Core  & 289   & 336   & 316   & 784   & 784   & 1034  & 453   & 1186  & 682  \\
          & Random & 325   & 403   & 368   & 862   & 860   & 1083  & 497   & 1242  & 841  \\
          & Edge  & 316   & 358   & 359   & 887   & 877   & 1126  & 520   & 1282  & 733  \\
    \bottomrule
    \end{tabular}
}
\end{table*}

\begin{figure}
    \centering
    \subfigure[ML-100K]{
        \includegraphics[width=0.315\textwidth]{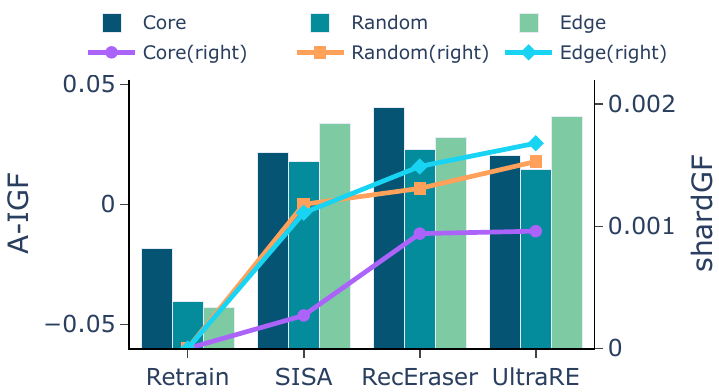}
    }
    \subfigure[ML-1M]{
        \includegraphics[width=0.315\textwidth]{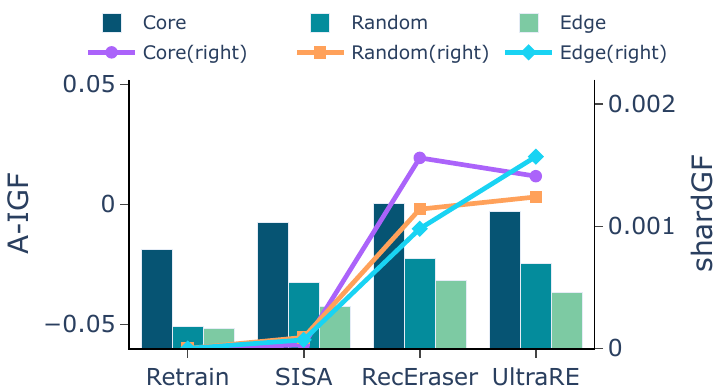}
    }
    \subfigure[ADM]{
        \includegraphics[width=0.315\textwidth]{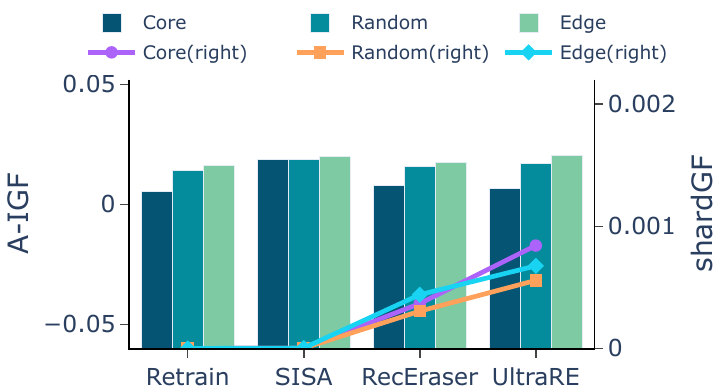}
    }
    \caption{Results in terms of recommendation fairness for exact recommendation unlearning methods on WMF, where A-IGF (approaching Retrain) and shardGF ($\downarrow$) evaluate the fairness of group-level and shard-level, respectively.
    }
    \label{fig:fair}
\end{figure}

\subsection{Unlearning Efficiency}
We report the unlearning times in Table~\ref{tab:time}. 
In general, SCIF is more efficient than EU methods. 
Among the EU methods, SISA saves more time compared to RecEraser and UltraRE, because it does not have the complex division and aggregation stage specific to the recommendation scenarios. 
Due to its design, UltraRE is slightly more efficient than RecEraser. 
Additionally, EU methods take less time to unlearn core users since they have a larger amount of interaction data.
This reduces the amount of data left for retraining.
On the contrary, SCIF requires more computations for influence estimation on core users, resulting in higher time costs compared to unlearning random or edge users.

\begin{figure}
    \centering
    \subfigure[Utility]{
        \includegraphics[width=0.23\textwidth]{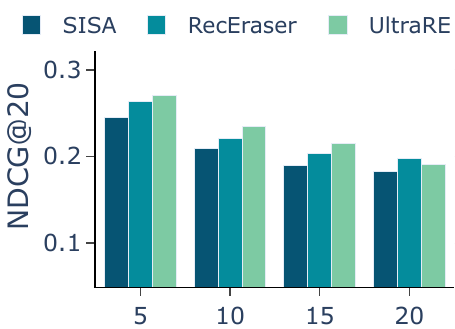}
    }
    \subfigure[Efficiency]{
        \includegraphics[width=0.23\textwidth]{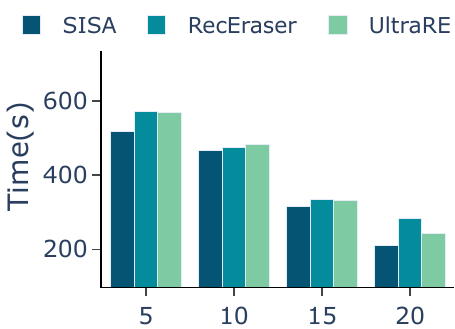}
    }
    \subfigure[Group-level Fairness]{
        \includegraphics[width=0.23\textwidth]{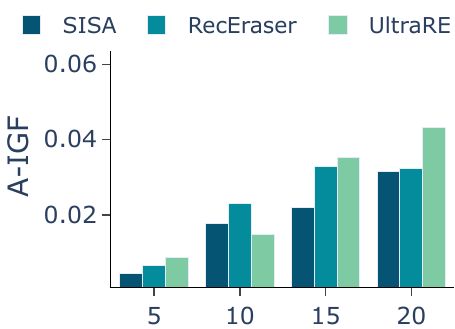}
    }
    \subfigure[Shard-level Fairness]{
        \includegraphics[width=0.23\textwidth]{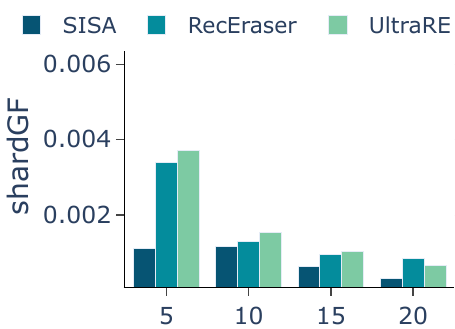}
    }
    
    \caption{
    Effect of shard number in terms of multiple aspects, i.e., recommendation utility ($\uparrow$), unlearning efficiency ($\downarrow$), group-level fairness (approaching Retrain), and shard-level fairness ($\downarrow$).
    }
    \label{fig:shard}
\end{figure}

\subsection{Recommendation Fairness}
We also report the recommendation fairness of AU and EU methods in Table~\ref{tab:au} and Figure~\ref{fig:fair}, respectively. 
For the \textit{group-level fairness} (A-IGF), compared to the AU method (SCIF), EU methods notably worsen unfairness, tending to favor active users.
This is primarily attributed to the division stage of EU methods, with this effect becoming more pronounced on larger datasets, i.e., ML-1M and ADM. 
Moreover, RecEraser and UltraRE, which group active users together instead of randomly, as done by SISA, exacerbate unfairness even further.
For the \textit{shard-level fairness} (shardGF), although to a lesser extent compared to group-level fairness, RecEraser and UltraRE also exacerbate unfairness.

\subsection{Effects of Shard Number}\label{sec:shard}
We report the effect of shard number in terms of multiple aspects in Figure~\ref{fig:shard}, using WMF on ML-100K.
\textit{Firstly}, as the number of shards increases, the unlearning efficiency improves, but the recommendation utility deteriorates, as confirmed by several previous studies~\citep{chen2022recommendation,li2023ultrare}.
\textit{Secondly}, the increased shard number further groups the active users into smaller shards, exacerbating the group-level fairness. At the same time, it reduces the discrepancy among all shards, diminishing the shard-level fairness.

\section{Conclusion}\label{sec:con}
In this paper, we present a comprehensive benchmark, \model{}, for recommendation unlearning methods, aiming to analyze and inspire further exploration into the deeper influence of recommendation unlearning. 
Specifically, \model{} covers four evaluation aspects, i.e., unlearning completeness, recommendation utility, unlearning efficiency, and recommendation fairness. 
Additionally, we investigate unlearning robustness across three unlearning sets, i.e., core data, edge data, and random data.
Through extensive experiments, we compare the performance of various recommendation unlearning methods using our proposed benchmark. 
Our experiments reveal that i) the division-aggregation design of the EU approach has dual implications. On one hand, it inherently achieves unlearning completeness. On the other hand, it compromises other evaluation aspects. and ii) The AU approach, which directly manipulates model parameters, outperforms the EU approach in all aspects except completeness, with less negative influence on model properties, e.g., fairness.
%

\paragraph{Limitation and Boarder Impact.} 
This paper proposes a benchmark for recommendation unlearning, comprising four evaluation aspects. This design also offers insights for other unlearning scenarios. 
Simultaneously, there is considerable room for improvement in the specific evaluation metrics within each aspect. 
%
%
Additionally, the AU approach appears to outperform the EU approach in all aspects except completeness. The trade-off between completeness and other aspects is an intriguing direction that is not discussed in this paper.

\acksection{}
This work was supported by the Fundamental Research Funds for the Central Universities 226-2024-00241. We thank all the anonymous reviewers for helpful feedback on early versions of this work.

\bibliographystyle{nips}
\bibliography{paper}

\begin{thebibliography}{52}
\providecommand{\natexlab}[1]{#1}
\providecommand{\url}[1]{\texttt{#1}}
\expandafter\ifx\csname urlstyle\endcsname\relax
  \providecommand{\doi}[1]{doi: #1}\else
  \providecommand{\doi}{doi: \begingroup \urlstyle{rm}\Url}\fi

\bibitem[Bourtoule et~al.(2021)Bourtoule, Chandrasekaran, Choquette-Choo, Jia, Travers, Zhang, Lie, and Papernot]{bourtoule2021machine}
Bourtoule, L., Chandrasekaran, V., Choquette-Choo, C.~A., Jia, H., Travers, A., Zhang, B., Lie, D., and Papernot, N.
\newblock Machine unlearning.
\newblock In \emph{2021 IEEE Symposium on Security and Privacy (SP)}, pp.\  141--159. IEEE, 2021.

\bibitem[Chen et~al.(2020)Chen, Zhang, Zhang, Liu, and Ma]{chen2020efficient}
Chen, C., Zhang, M., Zhang, Y., Liu, Y., and Ma, S.
\newblock Efficient neural matrix factorization without sampling for recommendation.
\newblock \emph{ACM Transactions on Information Systems (TOIS)}, 38\penalty0 (2):\penalty0 1--28, 2020.

\bibitem[Chen et~al.(2022)Chen, Sun, Zhang, and Ding]{chen2022recommendation}
Chen, C., Sun, F., Zhang, M., and Ding, B.
\newblock Recommendation unlearning.
\newblock In \emph{Proceedings of the ACM Web Conference 2022}, pp.\  2768--2777, 2022.

\bibitem[Chen et~al.(2024{\natexlab{a}})Chen, Zhang, Li, Wang, Qi, Xu, Zheng, and Yin]{chen2024post}
Chen, C., Zhang, Y., Li, Y., Wang, J., Qi, L., Xu, X., Zheng, X., and Yin, J.
\newblock Post-training attribute unlearning in recommender systems.
\newblock \emph{ACM Transactions on Information Systems}, 2024{\natexlab{a}}.

\bibitem[Chen \& Yang(2023)Chen and Yang]{chen2023unlearn}
Chen, J. and Yang, D.
\newblock Unlearn what you want to forget: Efficient unlearning for llms.
\newblock In \emph{Proceedings of the 2023 Conference on Empirical Methods in Natural Language Processing}, pp.\  12041--12052, 2023.

\bibitem[Chen et~al.(2024{\natexlab{b}})Chen, Yang, Xiong, Bai, Hu, Hao, Feng, Zhou, Wu, and Liu]{chen2024fast}
Chen, R., Yang, J., Xiong, H., Bai, J., Hu, T., Hao, J., Feng, Y., Zhou, J.~T., Wu, J., and Liu, Z.
\newblock Fast model debias with machine unlearning.
\newblock \emph{Advances in Neural Information Processing Systems}, 36, 2024{\natexlab{b}}.

\bibitem[Choi \& Na(2023)Choi and Na]{choi2023towards}
Choi, D. and Na, D.
\newblock Towards machine unlearning benchmarks: Forgetting the personal identities in facial recognition systems.
\newblock \emph{arXiv preprint arXiv:2311.02240}, 2023.

\bibitem[Eldan \& Russinovich(2023)Eldan and Russinovich]{eldan2023s}
Eldan, R. and Russinovich, M.
\newblock Who's harry potter? approximate unlearning in llms.
\newblock \emph{arXiv preprint arXiv:2310.02238}, 2023.

\bibitem[Fan et~al.(2024)Fan, Liu, Hero, and Liu]{fan2024challenging}
Fan, C., Liu, J., Hero, A., and Liu, S.
\newblock Challenging forgets: Unveiling the worst-case forget sets in machine unlearning.
\newblock \emph{arXiv preprint arXiv:2403.07362}, 2024.

\bibitem[Foster et~al.(2024)Foster, Schoepf, and Brintrup]{foster2024fast}
Foster, J., Schoepf, S., and Brintrup, A.
\newblock Fast machine unlearning without retraining through selective synaptic dampening.
\newblock In \emph{Proceedings of the AAAI Conference on Artificial Intelligence}, volume~38, pp.\  12043--12051, 2024.

\bibitem[Fredrikson et~al.(2015)Fredrikson, Jha, and Ristenpart]{fredrikson2015model}
Fredrikson, M., Jha, S., and Ristenpart, T.
\newblock Model inversion attacks that exploit confidence information and basic countermeasures.
\newblock In \emph{Proceedings of the 22nd ACM SIGSAC conference on computer and communications security}, pp.\  1322--1333, 2015.

\bibitem[Ganh{\"o}r et~al.(2022)Ganh{\"o}r, Penz, Rekabsaz, Lesota, and Schedl]{ganhor2022unlearning}
Ganh{\"o}r, C., Penz, D., Rekabsaz, N., Lesota, O., and Schedl, M.
\newblock Unlearning protected user attributes in recommendations with adversarial training.
\newblock In \emph{Proceedings of the 45th International ACM SIGIR Conference on Research and Development in Information Retrieval}, pp.\  2142--2147, 2022.

\bibitem[Graves et~al.(2021)Graves, Nagisetty, and Ganesh]{graves2021amnesiac}
Graves, L., Nagisetty, V., and Ganesh, V.
\newblock Amnesiac machine learning.
\newblock In \emph{Proceedings of the AAAI Conference on Artificial Intelligence}, volume~35, pp.\  11516--11524, 2021.

\bibitem[Gupta et~al.(2021)Gupta, Jung, Neel, Roth, Sharifi-Malvajerdi, and Waites]{gupta2021adaptive}
Gupta, V., Jung, C., Neel, S., Roth, A., Sharifi-Malvajerdi, S., and Waites, C.
\newblock Adaptive machine unlearning.
\newblock \emph{Advances in Neural Information Processing Systems}, 34:\penalty0 16319--16330, 2021.

\bibitem[Haveliwala(2002)]{haveliwala2002topic}
Haveliwala, T.~H.
\newblock Topic-sensitive pagerank.
\newblock In \emph{Proceedings of the 11th international conference on World Wide Web}, pp.\  517--526, 2002.

\bibitem[He et~al.(2017)He, Liao, Zhang, Nie, Hu, and Chua]{he2017neural}
He, X., Liao, L., Zhang, H., Nie, L., Hu, X., and Chua, T.-S.
\newblock Neural collaborative filtering.
\newblock In \emph{Proceedings of the 26th international conference on world wide web}, pp.\  173--182, 2017.

\bibitem[He et~al.(2020)He, Deng, Wang, Li, Zhang, and Wang]{he2020lightgcn}
He, X., Deng, K., Wang, X., Li, Y., Zhang, Y., and Wang, M.
\newblock Lightgcn: Simplifying and powering graph convolution network for recommendation.
\newblock In \emph{Proceedings of the 43rd International ACM SIGIR conference on research and development in Information Retrieval}, pp.\  639--648, 2020.

\bibitem[Hu et~al.(2024)Hu, Zhang, Xiao, Wang, Feng, and He]{hu2024exact}
Hu, Z., Zhang, Y., Xiao, M., Wang, W., Feng, F., and He, X.
\newblock Exact and efficient unlearning for large language model-based recommendation.
\newblock \emph{arXiv preprint arXiv:2404.10327}, 2024.

\bibitem[Information(2023)]{cal2023del}
Information, C.~L.
\newblock California senate bill 362, 2023.
\newblock URL \url{https://leginfo.legislature.ca.gov/faces/billNavClient.xhtml?bill_id=202120220SB362}.

\bibitem[Jin et~al.(2024)Jin, Cao, Wang, He, Yuan, Li, Chen, Liu, and Zhao]{jin2024rwku}
Jin, Z., Cao, P., Wang, C., He, Z., Yuan, H., Li, J., Chen, Y., Liu, K., and Zhao, J.
\newblock Rwku: Benchmarking real-world knowledge unlearning for large language models.
\newblock \emph{arXiv preprint arXiv:2406.10890}, 2024.

\bibitem[Kurmanji et~al.(2024)Kurmanji, Triantafillou, Hayes, and Triantafillou]{kurmanji2024towards}
Kurmanji, M., Triantafillou, P., Hayes, J., and Triantafillou, E.
\newblock Towards unbounded machine unlearning.
\newblock \emph{Advances in Neural Information Processing Systems}, 36, 2024.

\bibitem[Li et~al.(2024{\natexlab{a}})Li, Pan, Gopal, Yue, Berrios, Gatti, Li, Dombrowski, Goel, Phan, et~al.]{li2024wmdp}
Li, N., Pan, A., Gopal, A., Yue, S., Berrios, D., Gatti, A., Li, J.~D., Dombrowski, A.-K., Goel, S., Phan, L., et~al.
\newblock The wmdp benchmark: Measuring and reducing malicious use with unlearning.
\newblock \emph{arXiv preprint arXiv:2403.03218}, 2024{\natexlab{a}}.

\bibitem[Li et~al.(2012)Li, Ng, and Ye]{li2012har}
Li, X., Ng, M.~K., and Ye, Y.
\newblock Har: hub, authority and relevance scores in multi-relational data for query search.
\newblock In \emph{Proceedings of the 2012 SIAM International Conference on Data Mining}, pp.\  141--152. SIAM, 2012.

\bibitem[Li et~al.(2021)Li, Chen, Fu, Ge, and Zhang]{li2021user}
Li, Y., Chen, H., Fu, Z., Ge, Y., and Zhang, Y.
\newblock User-oriented fairness in recommendation.
\newblock In \emph{Proceedings of the web conference 2021}, pp.\  624--632, 2021.

\bibitem[Li et~al.(2023{\natexlab{a}})Li, Chen, Zhang, Liu, Lyu, Zheng, Meng, and Wang]{li2023ultrare}
Li, Y., Chen, C., Zhang, Y., Liu, W., Lyu, L., Zheng, X., Meng, D., and Wang, J.
\newblock Ultrare: Enhancing receraser for recommendation unlearning via error decomposition.
\newblock \emph{Advances in Neural Information Processing Systems}, 36, 2023{\natexlab{a}}.

\bibitem[Li et~al.(2023{\natexlab{b}})Li, Chen, Zheng, Zhang, Gong, Wang, and Chen]{li2023selective}
Li, Y., Chen, C., Zheng, X., Zhang, Y., Gong, B., Wang, J., and Chen, L.
\newblock Selective and collaborative influence function for efficient recommendation unlearning.
\newblock \emph{Expert Systems with Applications}, 234:\penalty0 121025, 2023{\natexlab{b}}.
\newblock ISSN 0957-4174.

\bibitem[Li et~al.(2023{\natexlab{c}})Li, Chen, Zheng, Zhang, Han, Meng, and Wang]{li2023making}
Li, Y., Chen, C., Zheng, X., Zhang, Y., Han, Z., Meng, D., and Wang, J.
\newblock Making users indistinguishable: Attribute-wise unlearning in recommender systems.
\newblock In \emph{Proceedings of the 31st ACM International Conference on Multimedia}, pp.\  984--994, 2023{\natexlab{c}}.

\bibitem[Li et~al.(2024{\natexlab{b}})Li, Feng, Chen, and Yang]{li2024survey}
Li, Y., Feng, X., Chen, C., and Yang, Q.
\newblock A survey on recommendation unlearning: Fundamentals, taxonomy, evaluation, and open questions.
\newblock \emph{arXiv preprint arXiv:2412.12836}, 2024{\natexlab{b}}.

\bibitem[Liu et~al.(2022)Liu, Fan, Chen, Liu, Ma, Wang, and Ma]{liu2022backdoor}
Liu, Y., Fan, M., Chen, C., Liu, X., Ma, Z., Wang, L., and Ma, J.
\newblock Backdoor defense with machine unlearning.
\newblock In \emph{IEEE INFOCOM 2022-IEEE conference on computer communications}, pp.\  280--289. IEEE, 2022.

\bibitem[Ma et~al.(2022)Ma, Liu, Liu, Liu, Ma, and Ren]{ma2022learn}
Ma, Z., Liu, Y., Liu, X., Liu, J., Ma, J., and Ren, K.
\newblock Learn to forget: Machine unlearning via neuron masking.
\newblock \emph{IEEE Transactions on Dependable and Secure Computing}, 2022.

\bibitem[Maini et~al.(2024)Maini, Feng, Schwarzschild, Lipton, and Kolter]{maini2024tofu}
Maini, P., Feng, Z., Schwarzschild, A., Lipton, Z.~C., and Kolter, J.~Z.
\newblock Tofu: A task of fictitious unlearning for llms.
\newblock \emph{arXiv preprint arXiv:2401.06121}, 2024.

\bibitem[Marchant et~al.(2022)Marchant, Rubinstein, and Alfeld]{marchant2022hard}
Marchant, N.~G., Rubinstein, B.~I., and Alfeld, S.
\newblock Hard to forget: Poisoning attacks on certified machine unlearning.
\newblock In \emph{Proceedings of the AAAI Conference on Artificial Intelligence}, volume~36, pp.\  7691--7700, 2022.

\bibitem[Mehta et~al.(2022)Mehta, Pal, Singh, and Ravi]{mehta2022deep}
Mehta, R., Pal, S., Singh, V., and Ravi, S.~N.
\newblock Deep unlearning via randomized conditionally independent hessians.
\newblock In \emph{Proceedings of the IEEE/CVF Conference on Computer Vision and Pattern Recognition}, pp.\  10422--10431, 2022.

\bibitem[Oesterling et~al.(2024)Oesterling, Ma, Calmon, and Lakkaraju]{oesterling2024fair}
Oesterling, A., Ma, J., Calmon, F., and Lakkaraju, H.
\newblock Fair machine unlearning: Data removal while mitigating disparities.
\newblock In \emph{International Conference on Artificial Intelligence and Statistics}, pp.\  3736--3744. PMLR, 2024.

\bibitem[Pardau(2018)]{pardau2018california}
Pardau, S.~L.
\newblock The california consumer privacy act: Towards a european-style privacy regime in the united states.
\newblock \emph{J. Tech. L. \& Pol'y}, 23:\penalty0 68, 2018.

\bibitem[Park et~al.(2019)Park, Kan, Dong, Zhao, and Faloutsos]{park2019estimating}
Park, N., Kan, A., Dong, X.~L., Zhao, T., and Faloutsos, C.
\newblock Estimating node importance in knowledge graphs using graph neural networks.
\newblock In \emph{Proceedings of the 25th ACM SIGKDD international conference on knowledge discovery \& data mining}, pp.\  596--606, 2019.

\bibitem[Rastegarpanah et~al.(2019)Rastegarpanah, Gummadi, and Crovella]{rastegarpanah2019fighting}
Rastegarpanah, B., Gummadi, K.~P., and Crovella, M.
\newblock Fighting fire with fire: Using antidote data to improve polarization and fairness of recommender systems.
\newblock In \emph{Proceedings of the twelfth ACM international conference on web search and data mining}, pp.\  231--239, 2019.

\bibitem[Rendle et~al.(2012)Rendle, Freudenthaler, Gantner, and Schmidt-Thieme]{rendle2012bpr}
Rendle, S., Freudenthaler, C., Gantner, Z., and Schmidt-Thieme, L.
\newblock Bpr: Bayesian personalized ranking from implicit feedback.
\newblock \emph{arXiv preprint arXiv:1205.2618}, 2012.

\bibitem[Sattigeri et~al.(2022)Sattigeri, Ghosh, Padhi, Dognin, and Varshney]{sattigeri2022fair}
Sattigeri, P., Ghosh, S., Padhi, I., Dognin, P., and Varshney, K.~R.
\newblock Fair infinitesimal jackknife: Mitigating the influence of biased training data points without refitting.
\newblock \emph{Advances in Neural Information Processing Systems}, 35:\penalty0 35894--35906, 2022.

\bibitem[Sekhari et~al.(2021)Sekhari, Acharya, Kamath, and Suresh]{sekhari2021remember}
Sekhari, A., Acharya, J., Kamath, G., and Suresh, A.~T.
\newblock Remember what you want to forget: Algorithms for machine unlearning.
\newblock \emph{Advances in Neural Information Processing Systems}, 34:\penalty0 18075--18086, 2021.

\bibitem[Thudi et~al.(2022)Thudi, Jia, Shumailov, and Papernot]{thudi2022necessity}
Thudi, A., Jia, H., Shumailov, I., and Papernot, N.
\newblock On the necessity of auditable algorithmic definitions for machine unlearning.
\newblock In \emph{31st USENIX Security Symposium (USENIX Security 22)}, pp.\  4007--4022, 2022.

\bibitem[Union(2018)]{euro2018gdpr}
Union, E.
\newblock General data protection regulation, 2018.
\newblock URL \url{https://gdpr-info.eu/}.

\bibitem[Wang et~al.(2024)Wang, Lin, Chen, Yang, Tang, Zhang, and Yu]{wang2024towards}
Wang, H., Lin, J., Chen, B., Yang, Y., Tang, R., Zhang, W., and Yu, Y.
\newblock Towards efficient and effective unlearning of large language models for recommendation.
\newblock \emph{arXiv preprint arXiv:2403.03536}, 2024.

\bibitem[Wang et~al.(2022)Wang, Guo, Xie, and Qi]{wang2022federated}
Wang, J., Guo, S., Xie, X., and Qi, H.
\newblock Federated unlearning via class-discriminative pruning.
\newblock In \emph{Proceedings of the ACM Web Conference 2022}, pp.\  622--632, 2022.

\bibitem[Wang et~al.(2019)Wang, He, Wang, Feng, and Chua]{wang2019neural}
Wang, X., He, X., Wang, M., Feng, F., and Chua, T.-S.
\newblock Neural graph collaborative filtering.
\newblock In \emph{Proceedings of the 42nd international ACM SIGIR conference on Research and development in Information Retrieval}, pp.\  165--174, 2019.

\bibitem[Wang et~al.(2023)Wang, Ma, Zhang, Liu, and Ma]{wang2023survey}
Wang, Y., Ma, W., Zhang, M., Liu, Y., and Ma, S.
\newblock A survey on the fairness of recommender systems.
\newblock \emph{ACM Transactions on Information Systems}, 41\penalty0 (3):\penalty0 1--43, 2023.

\bibitem[Wu et~al.(2022)Wu, Hashemi, and Srinivasa]{wu2022puma}
Wu, G., Hashemi, M., and Srinivasa, C.
\newblock Puma: Performance unchanged model augmentation for training data removal.
\newblock In \emph{Proceedings of the AAAI conference on artificial intelligence}, volume~36, pp.\  8675--8682, 2022.

\bibitem[Xin et~al.(2024)Xin, Yang, Zhao, Ren, Chen, Ma, and Ren]{xin2024effectiveness}
Xin, X., Yang, L., Zhao, Z., Ren, P., Chen, Z., Ma, J., and Ren, Z.
\newblock On the effectiveness of unlearning in session-based recommendation.
\newblock In \emph{Proceedings of the 17th ACM International Conference on Web Search and Data Mining}, pp.\  855--863, 2024.

\bibitem[Yan et~al.(2022)Yan, Li, Guo, Li, Li, and Lin]{yan2022arcane}
Yan, H., Li, X., Guo, Z., Li, H., Li, F., and Lin, X.
\newblock Arcane: An efficient architecture for exact machine unlearning.
\newblock In \emph{IJCAI}, volume~6, pp.\ ~19, 2022.

\bibitem[Ye \& Lu(2023)Ye and Lu]{ye2023sequence}
Ye, S. and Lu, J.
\newblock Sequence unlearning for sequential recommender systems.
\newblock In \emph{Australasian Joint Conference on Artificial Intelligence}, pp.\  403--415. Springer, 2023.

\bibitem[Zhang et~al.(2023)Zhang, Hu, Bai, Feng, Wu, Wang, and He]{zhang2023recommendation}
Zhang, Y., Hu, Z., Bai, Y., Feng, F., Wu, J., Wang, Q., and He, X.
\newblock Recommendation unlearning via influence function.
\newblock \emph{arXiv preprint arXiv:2307.02147}, 2023.

\bibitem[Zhang et~al.(2024)Zhang, Zhang, Yao, Jia, Liu, Liu, and Liu]{zhang2024unlearncanvas}
Zhang, Y., Zhang, Y., Yao, Y., Jia, J., Liu, J., Liu, X., and Liu, S.
\newblock Unlearncanvas: A stylized image dataset to benchmark machine unlearning for diffusion models.
\newblock \emph{arXiv preprint arXiv:2402.11846}, 2024.

\end{thebibliography}

\newpage
\section*{Checklist}

\begin{enumerate}
\item For all authors...
\begin{enumerate}
  \item Do the main claims made in the abstract and introduction accurately reflect the paper's contributions and scope?
    \answerYes{}
  \item Did you describe the limitations of your work?
    \answerYes{See Section~\ref{sec:con}.}  
  \item Did you discuss any potential negative societal impacts of your work?
    \answerYes{See Section~\ref{sec:con}.}  
  \item Have you read the ethics review guidelines and ensured that your paper conforms to them?
    \answerYes{} 
\end{enumerate}

\item If you are including theoretical results...
\begin{enumerate}
  \item Did you state the full set of assumptions of all theoretical results?
    \answerNA{}
    \item Did you include complete proofs of all theoretical results?
    \answerNA{}
\end{enumerate}

\item If you ran experiments (e.g. for benchmarks)...
\begin{enumerate}
  \item Did you include the code, data, and instructions needed to reproduce the main experimental results (either in the supplemental material or as a URL)?
    \answerYes{See Abstract.} 
  \item Did you specify all the training details (e.g., data splits, hyperparameters, how they were chosen)?
    \answerYes{See Section~\ref{sec:para}.} 
	\item Did you report error bars (e.g., with respect to the random seed after running experiments multiple times)?
    \answerNo{}
	\item Did you include the total amount of compute and the type of resources used (e.g., type of GPUs, internal cluster, or cloud provider)?
    \answerYes{See Section~\ref{sec:hard}.}
\end{enumerate}

\item If you are using existing assets (e.g., code, data, models) or curating/releasing new assets...
\begin{enumerate}
  \item If your work uses existing assets, did you cite the creators?
    \answerYes{See Section~\ref{sec:dataset}.}
  \item Did you mention the license of the assets?
    \answerNA{}
  \item Did you include any new assets either in the supplemental material or as a URL?
    \answerNo{}
  \item Did you discuss whether and how consent was obtained from people whose data you're using/curating?
    \answerNA{}
  \item Did you discuss whether the data you are using/curating contains personally identifiable information or offensive content?
    \answerNA{}
\end{enumerate}

\item If you used crowdsourcing or conducted research with human subjects...
\begin{enumerate}
  \item Did you include the full text of instructions given to participants and screenshots, if applicable?
   \answerNA{}
  \item Did you describe any potential participant risks, with links to Institutional Review Board (IRB) approvals, if applicable?
    \answerNA{}
  \item Did you include the estimated hourly wage paid to participants and the total amount spent on participant compensation?
    \answerNA{}
\end{enumerate}

\end{enumerate}

\newpage
\appendix
\onecolumn
\section{More Results}

\subsection{Performance Overview}
\begin{figure}[h]
    \subfigure[ML-100K]{
        \includegraphics[width=0.315\textwidth]{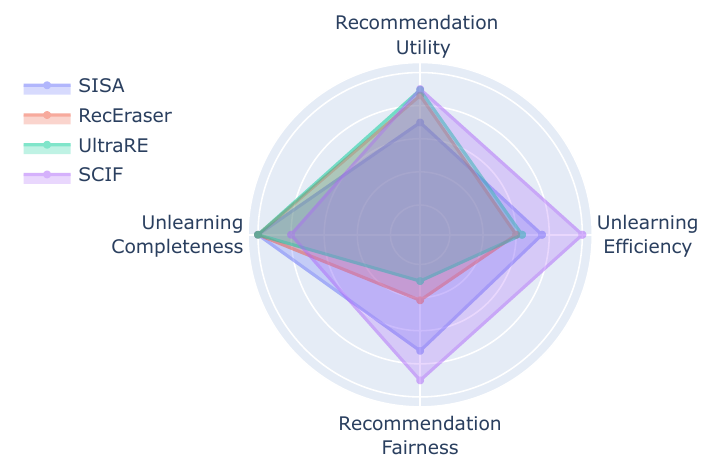}
    }
    \subfigure[ML-1M]{
        \includegraphics[width=0.315\textwidth]{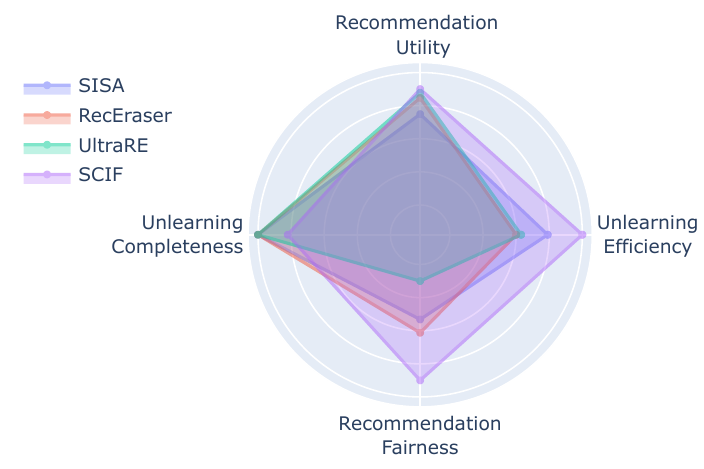}
    }
    \subfigure[ADM]{
        \includegraphics[width=0.315\textwidth]{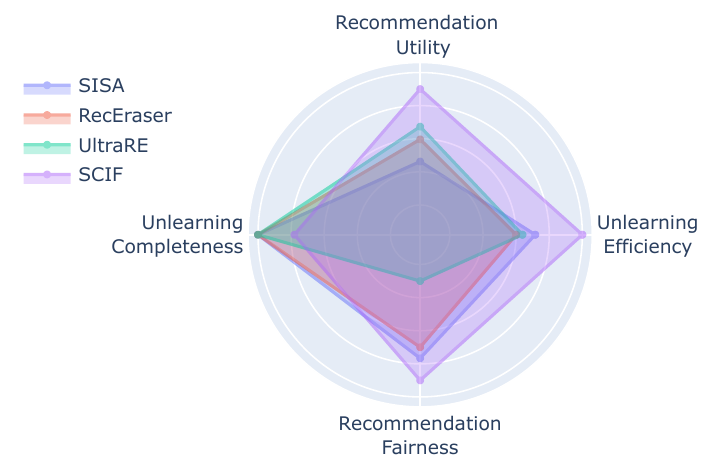}
    }
    \caption{
    A visualized evaluation overview of recommendation unlearning methods in four aspects ($\uparrow$), where the result is the normalized average outcome obtained across all models, using random data as the unlearning set. The recommendation fairness is measured by A-IGF (fairness between active and inactive users).
    }
    \label{fig:all_overview}
\end{figure}

We report a visualized overview of compared recommendation unlearning methods on each dataset in Figure~\ref{fig:all_overview}. The results are generally consistent with Figure~\ref{fig:overview}.

\subsection{Unlearning Completeness}\label{sec:comp_app}
We report the accuracy of MIO in Table~\ref{tab:aubpr}, where the recommendation model is BPR.
We omit the results for LightGCN as we encountered difficulties accurately computing the influence function of SCIF on ML-1M and ADM based on current hardware.

\begin{table*}[h]
\caption{Results in terms of unlearning completeness (MIO accuracy - approaching 0.5), recommendation utility (NDCG and HR $\uparrow$), and recommendation fairness (A-IGF - approaching Retrain) for the approximate recommendation unlearning method, where \texttt{Learn} denotes the results before unlearning.
Core, random, and edge respectively refer to the selection of the unlearning sets as core data, random data, and edge data.}
\label{tab:aubpr}
\centering
\resizebox{\textwidth}{!}{
    \begin{tabular}{cl|rrrr|rrrr|rrrr}
    \toprule
    \multicolumn{2}{c}{\multirow{2}[4]{*}{}} & \multicolumn{4}{c}{ML-100K}   & \multicolumn{4}{c}{ML-1M}     & \multicolumn{4}{c}{ADM} \\
\cmidrule{3-14}    \multicolumn{2}{c}{} & \multicolumn{1}{l}{NDCG@20} & \multicolumn{1}{l}{HR@20} & \multicolumn{1}{l}{MIO} & \multicolumn{1}{l}{A-IGF} & \multicolumn{1}{l}{NDCG@20} & \multicolumn{1}{l}{HR@20} & \multicolumn{1}{l}{MIO} & \multicolumn{1}{l}{A-IGF} & \multicolumn{1}{l}{NDCG@20} & \multicolumn{1}{l}{HR@20} & \multicolumn{1}{l}{MIO} & \multicolumn{1}{l}{A-IGF} \\
    \midrule
    \multicolumn{2}{c}{\texttt{Learn}} & 0.3195  & 0.3030  & 0.724  & -0.0246  & 0.2517 & 0.2306 & 0.744 & -0.0651 & 0.0251  & 0.0510  & 0.759 & 0.0194 \\
    \midrule
    \multirow{3}[2]{*}{Retrain} & Core  & 0.3111  & 0.3151  & 0.536  & -0.0217  & 0.2462  & 0.2279  & 0.549 & -0.0374  & 0.0246  & 0.0504  & 0.558 & 0.0066  \\
          & Random & 0.3003  & 0.3028  & 0.535  & -0.0153  & 0.2319  & 0.2162  & 0.550  & -0.0605  & 0.0203  & 0.0421  & 0.561 & 0.0187  \\
          & Edge  & 0.3043  & 0.2987  & 0.537  & -0.0175  & 0.2336  & 0.2118  & 0.552 & -0.0633  & 0.0203  & 0.0439  & 0.555 & 0.0191  \\
    \midrule
    \multirow{3}[2]{*}{SCIF} & Core  & 0.2392 & 0.2182 & 0.565  & -0.0116 & 0.1898 & 0.1636 & 0.572 & -0.0284 & 0.0171 & 0.0336 & 0.573 & 0.0096 \\
          & Random & 0.2768  & 0.2824  & 0.566  & -0.0144  & 0.2159  & 0.1886  & 0.576 & -0.0372  & 0.0189  & 0.0357  & 0.573 & 0.0110  \\
          & Edge  & 0.2871 & 0.2905 & 0.612  & -0.0167 & 0.2231 & 0.1942 & 0.635 & -0.0481 & 0.0200  & 0.0417 & 0.588 & 0.0132 \\
    \bottomrule
    \end{tabular}%
    }
\end{table*}%

\subsection{Recommendation Fairness}
\begin{figure}
    \centering
    \subfigure[ML-100K]{
        \includegraphics[width=0.315\textwidth]{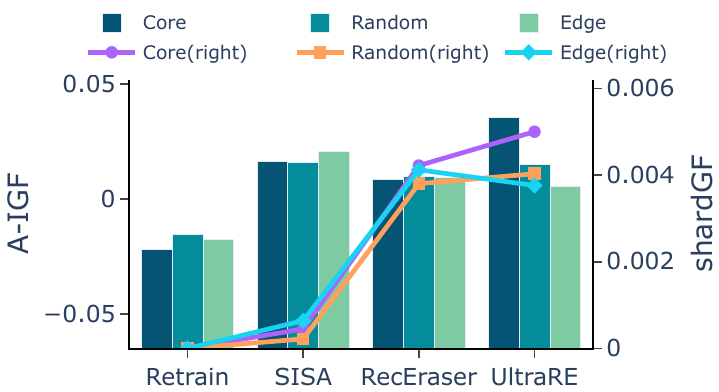}
    }
    \subfigure[ML-1M]{
        \includegraphics[width=0.315\textwidth]{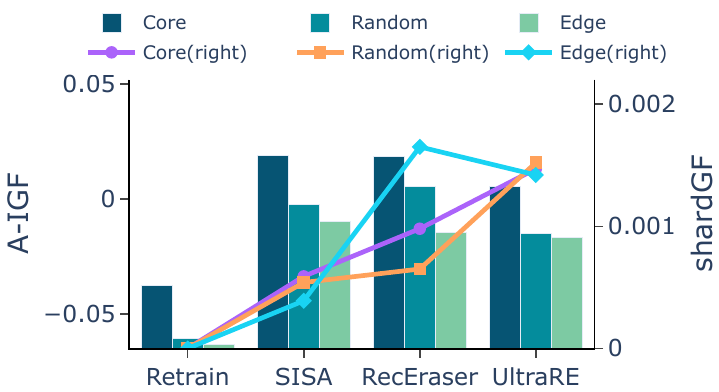}
    }
    \subfigure[ADM]{
        \includegraphics[width=0.315\textwidth]{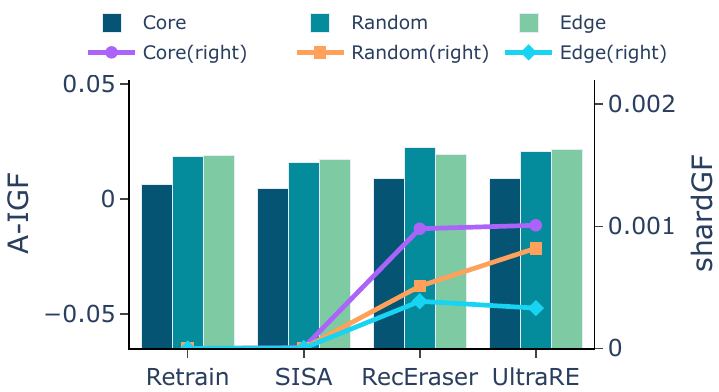}
    }
    \caption{Results in terms of recommendation fairness for exact recommendation unlearning methods on BPR, where A-IGF (approaching Retrain) and shardGF ($\downarrow$) evaluate the fairness of group-level and shard-level, respectively.
    }
    \label{fig:fair-bpr}
\end{figure}

\begin{figure}
    \centering
    \subfigure[ML-100K]{
        \includegraphics[width=0.315\textwidth]{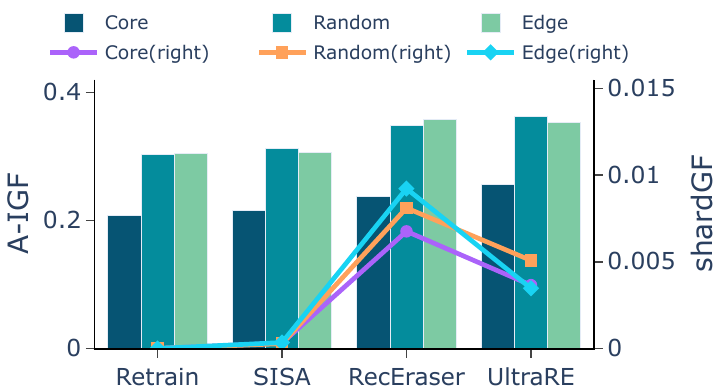}
    }
    \subfigure[ML-1M]{
        \includegraphics[width=0.315\textwidth]{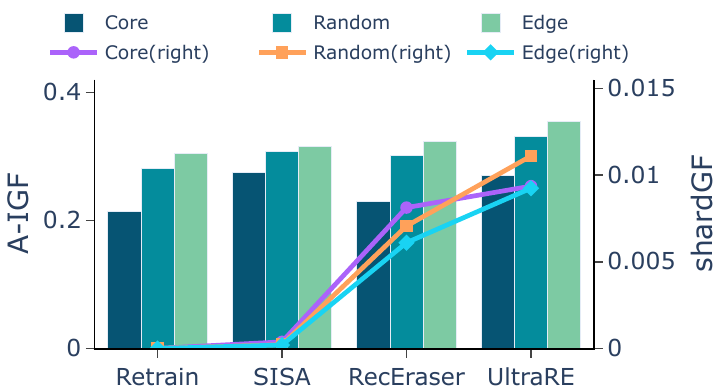}
    }
    \subfigure[ADM]{
        \includegraphics[width=0.315\textwidth]{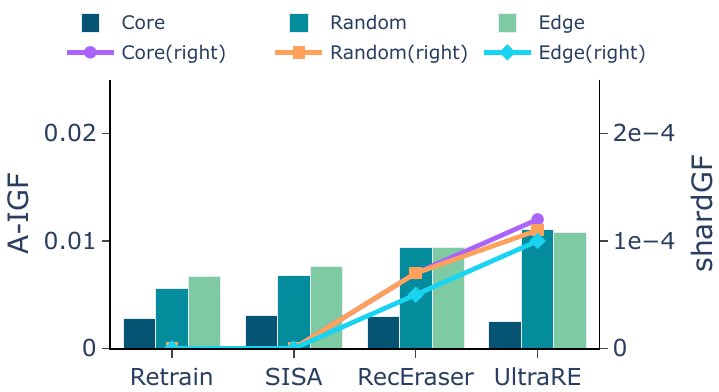}
    }
    \caption{Results in terms of recommendation fairness for exact recommendation unlearning methods on LightGCN, where A-IGF (approaching Retrain) and shardGF ($\downarrow$) evaluate the fairness of group-level and shard-level, respectively.
    }
    \label{fig:fair-light}
\end{figure}

We report the recommendation fairness of exact unlearning methods on each dataset using BPR and LightGCN recommendation models in Figures~\ref{fig:fair-bpr} and~\ref{fig:fair-light}, respectively.

We also report the grouping results of active and inactive users after applying three exact unlearning methods, i.e., SISA, RecEraser, UltraRE, on different datasets in Tables~\ref{tab:active_num_100k}, \ref{tab:active_num_1m}, and \ref{tab:active_num_adm}. 
On the one hand, SISA randomly distributes both types of users evenly across groups.
On the other hand, RecEraser and UltraRE tend to cluster active users into the same groups, which results in certain groups containing numerous active users while others have almost none. 
This clustering result explains why RecEraser and UltraRE tend to favor active users, as the concentration of active users in certain groups significantly increases their proportion compared to random distribution, leading to more effective learning but also more severe unfairness.

\subsection{Unlearning Ratio}\label{sec:ratio}
\begin{figure}
    \centering
    \subfigure[Utility]{
        \includegraphics[width=0.23\textwidth]{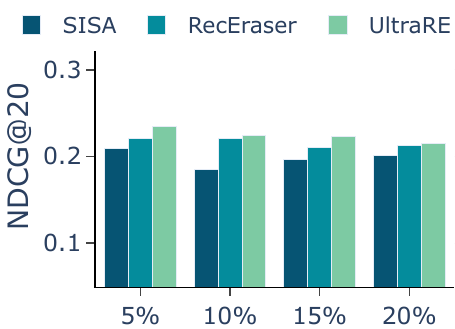}
    }
    \subfigure[Efficiency]{
        \includegraphics[width=0.23\textwidth]{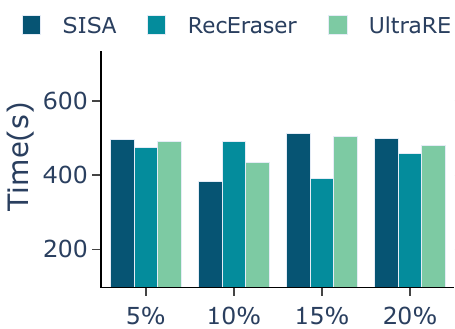}
    }
    \subfigure[Group-level Fairness]{
        \includegraphics[width=0.23\textwidth]{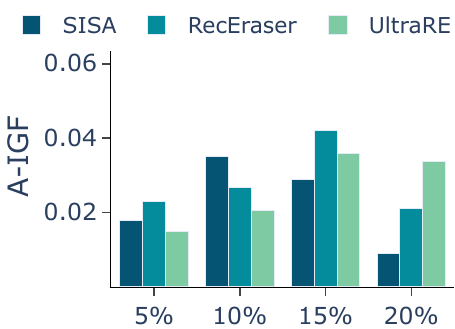}
    }
    \subfigure[Shard-level Fairness]{
        \includegraphics[width=0.23\textwidth]{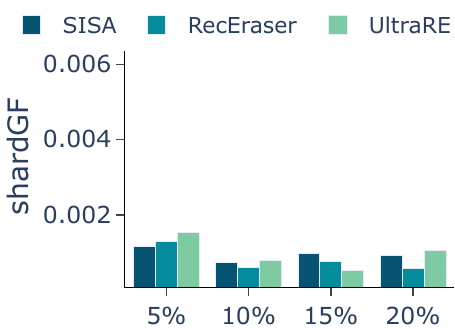}
    }
    \caption{
    Effect of unlearning ratio in terms of multiple aspects, i.e., recommendation utility ($\uparrow$), unlearning efficiency ($\downarrow$), group-level fairness (approaching Retrain), and shard-level fairness ($\downarrow$).
    }
    \label{fig:percent}
\end{figure}

We report the effect of unlearning data ratio in terms of multiple aspects in Figure~\ref{fig:percent}, using WMF on ML-100K.
We observe consistent results with previous studies~\citep{bourtoule2021machine, chen2022recommendation, li2023ultrare}. 
In general, as the ratio of unlearning data increases, the recommendation utility gradually decreases, along with a reduction in the unlearning time. Additionally, a larger unlearning ratio tends to lead to greater fairness.

\begin{table}[htbp]
  \centering
  \caption{Results of user distribution (active vs. inactive) in each shard on dataset ML-100K. The unlearning data ratio is set to 5\%.}
  \resizebox{\textwidth}{!}{
    \begin{tabular}{cl|rrrrrrrrrr}
    \toprule
    \multicolumn{2}{c}{\multirow{2}[4]{*}{ML-100K}} & \multicolumn{2}{c}{Group 1} & \multicolumn{2}{c}{Group 2} & \multicolumn{2}{c}{Group 3} & \multicolumn{2}{c}{Group 4} & \multicolumn{2}{c}{Group 5} \\
\cmidrule{3-12}    \multicolumn{2}{c}{} & \multicolumn{1}{l}{Active} & \multicolumn{1}{l}{Inactive} & \multicolumn{1}{l}{Active} & \multicolumn{1}{l}{Inactive} & \multicolumn{1}{l}{Active} & \multicolumn{1}{l}{Inactive} & \multicolumn{1}{l}{Active} & \multicolumn{1}{l}{Inactive} & \multicolumn{1}{l}{Active} & \multicolumn{1}{l}{Inactive} \\
    \midrule
    \multirow{3}[2]{*}{SISA} & Core  & 4     & 86    & 4     & 86    & 3     & 87    & 7     & 83    & 8     & 82 \\
          & Random & 4     & 86    & 8     & 82    & 2     & 88    & 3     & 87    & 6     & 84 \\
          & Edge  & 3     & 87    & 6     & 84    & 3     & 87    & 6     & 84    & 6     & 84 \\
    \midrule
    \multirow{3}[2]{*}{RecEraser} & Core  & 0     & 90    & 10    & 80    & 0     & 90    & 0     & 90    & 0     & 86 \\
          & Random & 0     & 90    & 1     & 89    & 1     & 89    & 0     & 90    & 0     & 90 \\
          & Edge  & 0     & 90    & 6     & 84    & 0     & 90    & 0     & 86    & 1     & 89 \\
    \midrule
    \multirow{3}[2]{*}{UltraRE} & Core  & 0     & 89    & 11    & 78    & 0     & 90    & 6     & 83    & 0     & 90 \\
          & Random & 0     & 90    & 3     & 86    & 3     & 87    & 1     & 89    & 0     & 89 \\
          & Edge  & 0     & 89    & 15    & 75    & 1     & 88    & 2     & 87    & 7     & 83 \\
    \midrule
    \multicolumn{2}{c}{\multirow{2}[4]{*}{ML-100K}} & \multicolumn{2}{c}{Group 6} & \multicolumn{2}{c}{Group 7} & \multicolumn{2}{c}{Group 8} & \multicolumn{2}{c}{Group 9} & \multicolumn{2}{c}{Group 10} \\
\cmidrule{3-12}    \multicolumn{2}{c}{} & \multicolumn{1}{l}{Active} & \multicolumn{1}{l}{Inactive} & \multicolumn{1}{l}{Active} & \multicolumn{1}{l}{Inactive} & \multicolumn{1}{l}{Active} & \multicolumn{1}{l}{Inactive} & \multicolumn{1}{l}{Active} & \multicolumn{1}{l}{Inactive} & \multicolumn{1}{l}{Active} & \multicolumn{1}{l}{Inactive} \\
    \midrule
    \multirow{3}[2]{*}{SISA} & Core  & 1     & 89    & 3     & 86    & 3     & 86    & 6     & 83    & 5     & 84 \\
          & Random & 5     & 85    & 4     & 85    & 1     & 88    & 5     & 84    & 6     & 83 \\
          & Edge  & 8     & 82    & 2     & 87    & 1     & 88    & 7     & 82    & 2     & 87 \\
    \midrule
    \multirow{3}[2]{*}{RecEraser} & Core  & 0     & 90    & 0     & 90    & 28    & 62    & 0     & 90    & 6     & 84 \\
          & Random & 6     & 84    & 9     & 81    & 0     & 90    & 27    & 63    & 0     & 86 \\
          & Edge  & 0     & 90    & 9     & 81    & 0     & 90    & 27    & 63    & 1     & 89 \\
    \midrule
    \multirow{3}[2]{*}{UltraRE} & Core  & 7     & 83    & 10    & 80    & 1     & 89    & 0     & 89    & 9     & 81 \\
          & Random & 3     & 87    & 1     & 88    & 0     & 89    & 18    & 72    & 15    & 75 \\
          & Edge  & 0     & 90    & 1     & 89    & 0     & 90    & 7     & 82    & 11    & 79 \\
    \bottomrule
    \end{tabular}%
    }
  \label{tab:active_num_100k}%
\end{table}

\begin{table}[htbp]
  \centering
  \caption{Results of user distribution (active vs. inactive) in each shard on dataset ML-1M. The unlearning data ratio is set to 5\%.}
  \resizebox{\textwidth}{!}{
    \begin{tabular}{cl|rrrrrrrrrr}
    \toprule
    \multicolumn{2}{c}{\multirow{2}[4]{*}{ML-1M}} & \multicolumn{2}{c}{Group 1} & \multicolumn{2}{c}{Group 2} & \multicolumn{2}{c}{Group 3} & \multicolumn{2}{c}{Group 4} & \multicolumn{2}{c}{Group 5} \\
\cmidrule{3-12}    \multicolumn{2}{c}{} & \multicolumn{1}{l}{Active} & \multicolumn{1}{l}{Inactive} & \multicolumn{1}{l}{Active} & \multicolumn{1}{l}{Inactive} & \multicolumn{1}{l}{Active} & \multicolumn{1}{l}{Inactive} & \multicolumn{1}{l}{Active} & \multicolumn{1}{l}{Inactive} & \multicolumn{1}{l}{Active} & \multicolumn{1}{l}{Inactive} \\
    \midrule
    \multirow{3}[2]{*}{SISA} & Core  & 28    & 546   & 31    & 543   & 26    & 548   & 30    & 544   & 32    & 542 \\
          & Random & 25    & 549   & 34    & 540   & 30    & 544   & 23    & 551   & 35    & 539 \\
          & Edge  & 36    & 538   & 20    & 554   & 24    & 550   & 32    & 542   & 27    & 547 \\
    \midrule
    \multirow{3}[2]{*}{RecEraser} & Core  & 44    & 530   & 52    & 522   & 0     & 572   & 20    & 554   & 5     & 569 \\
          & Random & 2     & 570   & 79    & 495   & 44    & 530   & 40    & 534   & 5     & 569 \\
          & Edge  & 10    & 564   & 41    & 533   & 74    & 500   & 31    & 543   & 0     & 574 \\
    \midrule
    \multirow{3}[2]{*}{UltraRE} & Core  & 0     & 573   & 5     & 569   & 11    & 563   & 12    & 562   & 33    & 541 \\
          & Random & 44    & 530   & 6     & 567   & 7     & 567   & 5     & 569   & 13    & 561 \\
          & Edge  & 8     & 566   & 9     & 564   & 11    & 563   & 4     & 569   & 23    & 550 \\
    \midrule
    \multicolumn{2}{c}{\multirow{2}[4]{*}{ML-1M}} & \multicolumn{2}{c}{group6} & \multicolumn{2}{c}{Group 7} & \multicolumn{2}{c}{Group 8} & \multicolumn{2}{c}{Group 9} & \multicolumn{2}{c}{Group 10} \\
\cmidrule{3-12}    \multicolumn{2}{c}{} & \multicolumn{1}{l}{Active} & \multicolumn{1}{l}{Inactive} & \multicolumn{1}{l}{Active} & \multicolumn{1}{l}{Inactive} & \multicolumn{1}{l}{Active} & \multicolumn{1}{l}{Inactive} & \multicolumn{1}{l}{Active} & \multicolumn{1}{l}{Inactive} & \multicolumn{1}{l}{Active} & \multicolumn{1}{l}{Inactive} \\
    \midrule
    \multirow{3}[2]{*}{SISA} & Core  & 23    & 551   & 25    & 549   & 33    & 541   & 28    & 545   & 30    & 543 \\
          & Random & 28    & 546   & 38    & 536   & 22    & 552   & 30    & 543   & 21    & 552 \\
          & Edge  & 27    & 547   & 39    & 535   & 33    & 541   & 26    & 547   & 22    & 551 \\
    \midrule
    \multirow{3}[2]{*}{RecEraser} & Core  & 64    & 510   & 1     & 573   & 38    & 536   & 35    & 539   & 27    & 547 \\
          & Random & 32    & 542   & 1     & 573   & 61    & 513   & 2     & 572   & 20    & 554 \\
          & Edge  & 24    & 550   & 2     & 572   & 91    & 483   & 4     & 568   & 9     & 565 \\
    \midrule
    \multirow{3}[2]{*}{UltraRE} & Core  & 44    & 530   & 27    & 547   & 18    & 556   & 32    & 541   & 104   & 470 \\
          & Random & 3     & 571   & 0     & 574   & 14    & 559   & 7     & 567   & 187   & 387 \\
          & Edge  & 0     & 575   & 49    & 525   & 26    & 548   & 155   & 419   & 1     & 573 \\
    \bottomrule
    \end{tabular}%
    }
  \label{tab:active_num_1m}%
\end{table}

\begin{table}[htbp]
  \centering
  \caption{Results of user distribution (active vs inactive) in each shard on dataset ADM. The unlearning data ratio is set to 5\%.}
  \resizebox{\textwidth}{!}{
    \begin{tabular}{cl|rrrrrrrrrr}
    \toprule
    \multicolumn{2}{c}{\multirow{2}[4]{*}{ADM}} & \multicolumn{2}{c}{Group 1} & \multicolumn{2}{c}{Group 2} & \multicolumn{2}{c}{Group 3} & \multicolumn{2}{c}{Group 4} & \multicolumn{2}{c}{Group 5} \\
\cmidrule{3-12}    \multicolumn{2}{c}{} & \multicolumn{1}{l}{Active} & \multicolumn{1}{l}{Inactive} & \multicolumn{1}{l}{Active} & \multicolumn{1}{l}{Inactive} & \multicolumn{1}{l}{Active} & \multicolumn{1}{l}{Inactive} & \multicolumn{1}{l}{Active} & \multicolumn{1}{l}{Inactive} & \multicolumn{1}{l}{Active} & \multicolumn{1}{l}{Inactive} \\
    \midrule
    \multirow{3}[2]{*}{SISA} & Core  & 96    & 2078  & 113   & 2061  & 126   & 2048  & 117   & 2057  & 106   & 2068 \\
          & Random & 108   & 2066  & 112   & 2062  & 95    & 2079  & 110   & 2064  & 84    & 2090 \\
          & Edge  & 112   & 2062  & 105   & 2069  & 100   & 2074  & 120   & 2054  & 106   & 2068 \\
    \midrule
    \multirow{3}[2]{*}{RecEraser} & Core  & 429   & 1745  & 0     & 2174  & 8     & 2166  & 0     & 2174  & 0     & 2169 \\
          & Random & 0     & 2169  & 0     & 2174  & 453   & 1721  & 159   & 2015  & 84    & 2090 \\
          & Edge  & 149   & 2025  & 84    & 2090  & 379   & 1795  & 7     & 2167  & 0     & 2169 \\
    \midrule
    \multirow{3}[2]{*}{UltraRE} & Core  & 91    & 2083  & 160   & 2013  & 65    & 2108  & 65    & 2109  & 88    & 2086 \\
          & Random & 41    & 2132  & 80    & 2094  & 361   & 1813  & 81    & 2093  & 56    & 2117 \\
          & Edge  & 82    & 2092  & 11    & 2162  & 201   & 1972  & 53    & 2120  & 330   & 1844 \\
    \midrule
    \multicolumn{2}{c}{\multirow{2}[4]{*}{ADM}} & \multicolumn{2}{c}{Group 6} & \multicolumn{2}{c}{Group 7} & \multicolumn{2}{c}{Group 8} & \multicolumn{2}{c}{Group 9} & \multicolumn{2}{c}{Group 10} \\
\cmidrule{3-12}    \multicolumn{2}{c}{} & \multicolumn{1}{l}{Active} & \multicolumn{1}{l}{Inactive} & \multicolumn{1}{l}{Active} & \multicolumn{1}{l}{Inactive} & \multicolumn{1}{l}{Active} & \multicolumn{1}{l}{Inactive} & \multicolumn{1}{l}{Active} & \multicolumn{1}{l}{Inactive} & \multicolumn{1}{l}{Active} & \multicolumn{1}{l}{Inactive} \\
    \midrule
    \multirow{3}[2]{*}{SISA} & Core  & 98    & 2075  & 112   & 2061  & 99    & 2074  & 115   & 2058  & 104   & 2069 \\
          & Random & 119   & 2054  & 114   & 2059  & 135   & 2038  & 116   & 2057  & 93    & 2080 \\
          & Edge  & 109   & 2064  & 106   & 2067  & 111   & 2062  & 102   & 2071  & 115   & 2058 \\
    \midrule
    \multirow{3}[2]{*}{RecEraser} & Core  & 1     & 2173  & 97    & 2077  & 388   & 1786  & 121   & 2053  & 42    & 2132 \\
          & Random & 9     & 2165  & 0     & 2174  & 0     & 2174  & 4     & 2170  & 377   & 1797 \\
          & Edge  & 456   & 1718  & 0     & 2174  & 0     & 2174  & 10    & 2164  & 1     & 2173 \\
    \midrule
    \multirow{3}[2]{*}{UltraRE} & Core  & 65    & 2109  & 173   & 2001  & 147   & 2026  & 123   & 2051  & 109   & 2063 \\
          & Random & 200   & 1973  & 50    & 2124  & 137   & 2036  & 48    & 2125  & 32    & 2142 \\
          & Edge  & 82    & 2091  & 58    & 2116  & 121   & 2052  & 55    & 2119  & 93    & 2081 \\
    \bottomrule
    \end{tabular}%
    }
  \label{tab:active_num_adm}%
\end{table}

\end{document}